\documentclass[aps,prl,twocolumn,groupedaddress]{revtex4}  
\usepackage{bm,graphicx,mathrsfs}
\usepackage{graphicx}
\usepackage{epsfig}
\usepackage{amsmath,bbm}
\usepackage{amsfonts,amssymb}
\usepackage{times}
\usepackage{lipsum}

\usepackage{latexsym}
\usepackage{amssymb}
\usepackage{amsmath}
\usepackage[colorlinks = true]{hyperref}

\usepackage{color}
\usepackage{verbatim}

\begin{document}

\title{Exploiting dynamic quantum circuits in a quantum algorithm with superconducting qubits}

\author{A. D. C\'orcoles}
\author{Maika Takita}
\author{Ken Inoue}
\author{Scott Lekuch}
\author{Zlatko K. Minev}
\author{Jerry M. Chow}
\author{Jay M. Gambetta}
\affiliation{IBM Quantum, IBM T.J.  Watson  Research  Center,  Yorktown  Heights,  NY 10598,  USA}
\date{\today}
\begin{abstract}
		The execution of quantum circuits on real systems has largely been limited to those which are simply time-ordered sequences of unitary operations followed by a projective measurement. As hardware platforms for quantum computing continue to mature in size and capability,  it is imperative to enable quantum circuits beyond their conventional construction. Here we break into the realm of dynamic quantum circuits on a superconducting-based quantum system. Dynamic quantum circuits involve not only the evolution of the quantum state throughout the computation, but also periodic measurements of a subset of qubits mid-circuit and concurrent processing of the resulting classical information within timescales shorter than the execution times of the circuits. Using noisy quantum hardware, we explore one of the most fundamental quantum algorithms, quantum phase estimation, in its adaptive version, which exploits dynamic circuits, and compare the results to a non-adaptive implementation of the same algorithm. We demonstrate that the version of real-time quantum computing with dynamic circuits can offer a substantial and tangible advantage when noise and latency are sufficiently low in the system, opening the door to a new realm of available algorithms on real quantum systems. 
\end{abstract}
	
\maketitle

The evolution of quantum information processing in real quantum systems has taken remarkable leaps in recent years, transcending from laboratory demonstrations to systems with reliability and performance suitable for cloud-based research access \cite{IQX,Honeywell}. Not long ago, experimental efforts largely focused on understanding the components that make up a quantum system, from understanding the limits to qubit coherence \cite{Allcock11, Gambetta17, Wang15}, state control \cite{Harty14, Ospelkaus11, Chow09, Lucero10}, and readout \cite{Riste12, Walter17} with a significant amount of focus on developing improved two-qubit entangling gates \cite{Chow11, Barends14, Ballance16} while lowering crosstalk \cite{Gambetta12,Takita16,McKay20}. Today there is a significant shift emerging towards implementing quantum circuits for exploring interesting new algorithms \cite{havlicek19,Kandala17} and as a tool for benchmarking the quality of a quantum computer \cite{Cross19}.

A quantum circuit is a computational routine consisting of coherent quantum operations on quantum data, such as qubits, and concurrent real-time classical computation. Most early experiments with qubits involve quantum circuits that are simple in nature and consist of an ordered sequence of resets for qubit initialization, followed by quantum gates, and measurements. The simplicity of these circuits lies in the fact that they do not require any classical logic to be performed in the coherence time of the qubits. While this class is sufficient to implement the circuit model of quantum computing for practical implementations of quantum computing it is not enough for dynamic circuits that include quantum error correction, quantum teleportation, and iterative phase estimation, and even the measurement model of quantum computing. Dynamic circuits are circuits in which future states depend on outcomes of measurements that happen during the circuit.

Today, we are beginning to see emerging experiments where this classical \emph{real-time} logic is built into the circuits. Examples include mid-circuit measurement \cite{Griffiths96,Minev19}, mid-circuit measurement \cite{Riste12, Ofek16, Andersen19}, and demonstrations requiring low branching complexity such as quantum state and gate teleportation \cite{Barrett04, Riebe04, Steffen13, Chou18}, state injection \cite{Ryan17}, and initial demonstrations of dynamic quantum error correction \cite{Reinhold20}. In these examples, the classical logic either requires no state information or only requires small amounts of information within simple algorithms. Therefore, these dynamic circuits can be implemented simply even for reasonably large systems. Here, we  demonstrate a more demanding class of dynamic circuits with requirements for accurate quantum operations, measurements and resets as well as efficient handling of a large throughput of classical information in a time scale commensurate with the system's coherence. One of the protocols that better embodies the critical need for synergy between classical and quantum hardware in complex dynamic circuits is coincidentally an efficient version of the most important building piece of any quantum algorithm with a known exponential speed-up: the Quantum Phase Estimation (QPE) algorithm~\cite{nielsen00,Cleve98}. With this demonstration we unveil a hitherto experimentally unexplored regime in quantum information processing. As quantum systems get increasingly accurate, longer lived, and faster queried, it is important to consider pathways for the processing of their classical outputs that does not limit the capability of the quantum system to compute, neither in time nor in breadth of resources. Our work signifies a first step towards dynamic circuits with non-trivial complexity.

QPE is a family of algorithms whose object is the efficient eigenvalue sampling of a Hamiltonian. Some flavors of QPE include the standard Quantum Fourier Transform (QFT) version~\cite{nielsen00, Cleve98}, approximate QFT methods~\cite{Barenco96}, semiclassical QFT \cite{Griffiths96}, Kitaev QPE~\cite{Kitaev95}, Iterative Phase Estimation (IPE)~\cite{Childs2000,Knill07,Dobsicek07}, Heisenberg-limited QPE~\cite{Higgins09}, and Bayesian QPE \cite{Wiebe16,Paesani17}. 
\begin{figure*}[t!]
	\centering
	\includegraphics[width=\textwidth]{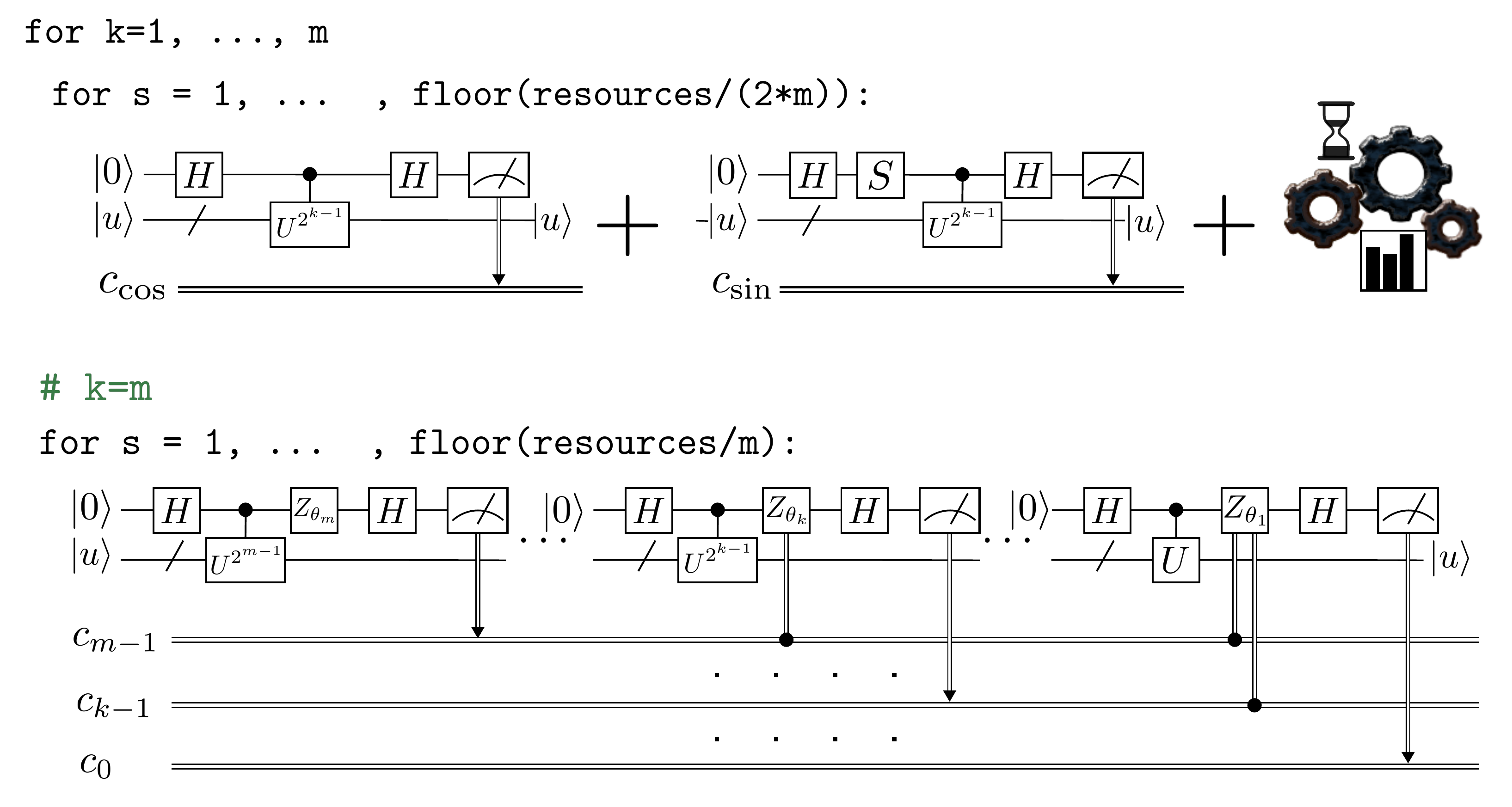}
	\caption{\label{Fig1} Circuit diagrams for the two approaches of QPE studied in this work: Kitaev's QPE (top) and IPE (bottom). The Kitaev circuits do not need previous knowledge of other bits in the phase at the expense of running and measuring one extra circuit per bit and needing additional near-time computing resources. The iterative version uses a $Z-$rotation conditional on all the previous measured bits before measuring the auxiliary qubit in the $X$-basis.}
\end{figure*}
QPE has been studied theoretically in noisy systems before~\cite{O_Brien_2019} and alternative methods have been proposed that accelerate the learning at the cost of -potentially exponential- classical post-processing of the outputs~\cite{Svore14}. In this work, however, we want to explore QPE in the presence of limited computational resources, while also assessing the impact of real-time classical operations on quantum systems and the exploitation of dynamic circuits. With the advent of cloud quantum systems for quantum computing~\cite{IQX,Honeywell}, the community has witnessed a shift on the user profile, from almost exclusively quantum information processing researchers, to a rather diverse set which includes physicists, computer scientists, and educators. The increased availability and popularity of state-of-the-art quantum computing systems over the cloud has started to put to the test the capacity of such systems to process and satisfy all users' demands in a timely manner. Under this scenario, one could argue that each measurement or sampling of a quantum system is a coveted and valuable resource.  From there, the question that naturally arises is: what is the minimum number of resources required to achieve a given accuracy in the answer to a quantum algorithm in a noisy system?

QPE is concerned with the problem of estimating an eigenvalue $\phi$ given the corresponding eigenstate $|u\rangle$ for the unitary operator $U$, with
\begin{equation}
U|u\rangle = e^{i\phi}|u\rangle
\end{equation}
We can use a quantum register to encode the eigenstate $|u\rangle$, and an auxiliary system, which we can call the pointer, to extract the phase. The simplest method for solving this problem is to run the left-hand side circuit in the top panel of Fig. \ref{Fig1} with $k=1$. By preparing the pointer qubit in the $|+\rangle$ state and using it as the control in a controlled-$U$ operation, the phase $\phi$ can be transferred from the eigenstate register to the pointer. If we define $\varphi = \phi/2\pi$, sampling the distribution obtained when measuring the pointer in the $x-$basis will give us the $m-$bit approximation $\tilde{\varphi}=\sum_{k=1}^m \varphi_k / 2^k = .\varphi_1 \varphi_2 \dots \varphi_m$ with $\varphi_k \in \{0,1\}$, with $|\tilde{\varphi} - \varphi| \leq 1/2^{m+1}$, where we use the standard dot notation for a binary expansion. In order to estimate the sampling resources needed for $1/2^{m+1}$ accuracy, we can consider the Hoeffding's inequality
\begin{equation}
\mathrm{Pr}(|\tilde{\varphi} - \varphi| \geq \epsilon) \leq 2e^{-2\epsilon^2 s}
\end{equation}
where $\epsilon = 1/2^{m+1}$, and $s$ is the number of circuit samplings. We can thus approximate the phase within accuracy $\epsilon$ with probability $1-\delta$ in $\mathcal{O} \Big(\frac{ \log ( 1 /\delta)}{\epsilon^2} \Big)$ samples, which is exponentially costly in $m$.

Kitaev's original approach to QPE uses the two families of circuits depicted in Fig. \ref{Fig1} (top). From these circuits, we can obtain an approximation to the quantities $\alpha_k = 2^{k-1} \tilde{\varphi}$ as
\begin{equation*}
	\begin{split}
	\cos(2\pi \alpha_k) = 2P(0|k) -1 \\
	\sin(2\pi \alpha_k) = 1 - 2P(0|k)
	\end{split}    
\end{equation*}
where the cosine refers to the left-hand circuit and the sine refers to the right-hand circuit, $P(0|k)$ is the probability of measuring the pointer in 0 for the circuit corresponding to bit $k$, and $\alpha_k$ shifts the bits in $\varphi$ to the left by $k$ positions. The two types of circuits are needed to lift the sign uncertainty in $\alpha_k$. These phase shifts can then be used within a classical iterative algorithm to reconstruct our approximation to the phase $\tilde{\varphi}$ \cite{Supp}, which relies on near-time computing for a final algorithmic answer.

The key aspect of Kitaev's approach is the extraction of the phase shifts $\alpha_k$ one at a time by using different powers of the unitary $U$. There is a potentially exponential cost associated to the implementation of these controlled-$U^{2^{k-1}}$ operations, but for the scope of this work we will assume each power of $U$ is equally efficient to implement. Provided we can estimate each of the $\alpha_k$ with constant precision, we can estimate the phase to accuracy $1/2^{m+1}$ efficiently using Kitaev's algorithm. To see this we simply apply the Chernoff bound to \emph{each} of the Hadamard tests in the top panel of Fig. \ref{Fig1}, assuming each probability of estimating $\alpha_k$ within constant accuracy is bounded by $\bar{\delta}$, and note that $2m$ estimates are required in this case. It can then be shown that we can obtain the phase within accuracy $1/2^{m+1}$ with probability $1-\delta$, where $\delta = 2m\bar{\delta}$, in $\mathcal{O} (m \log (m/\delta))$ samples, which is no longer exponential in $m$. This is a natural consequence of all the phase shifts being obtained with the same accuracy by virtue of using powers of the unitary $U$.

One could do better than this approach by exploiting dynamic circuits. By running the single dynamic quantum circuit in Fig. \ref{Fig1} (bottom), where measurements and interspersed throughout the circuit and classical information is processed within the duration of it, we can economize in the number of resources spent and eliminate the need for the classical post-processing. This is the IPE approach, where each of the measurements of the pointer following a controlled-$U^{2^{k-1}}$ provides the \emph{kth-}bit $\varphi_k$ directly. The IPE algorithm thus builds the phase from least to most significant bit, and is adaptive in  nature, meaning that the exact configuration of the circuit gates depends on the outcomes observed in the measurements performed throughout the circuit itself, via the argument $\theta_k/2\pi = - .0\varphi_{k+1} \varphi_{k+2} \dots \varphi_m$, where $\theta_m = 0$. In the absence of noise, it can be shown \cite{Kaye07} that a single measurement of the pointer for each sub-circuit gives an $m-$bit approximation to the phase with probability higher than $8/\pi^2$. For superconducting qubits, and in order to minimize impact from qubit decoherence, the successful implementation of this algorithm, as of any other type of dynamic circuit, requires classical control electronics capable of handling the information arriving from the quantum processor and acting upon it at a rate as high as possible. 

	\begin{figure}[t!]
		\centering
		\includegraphics[width=\columnwidth]{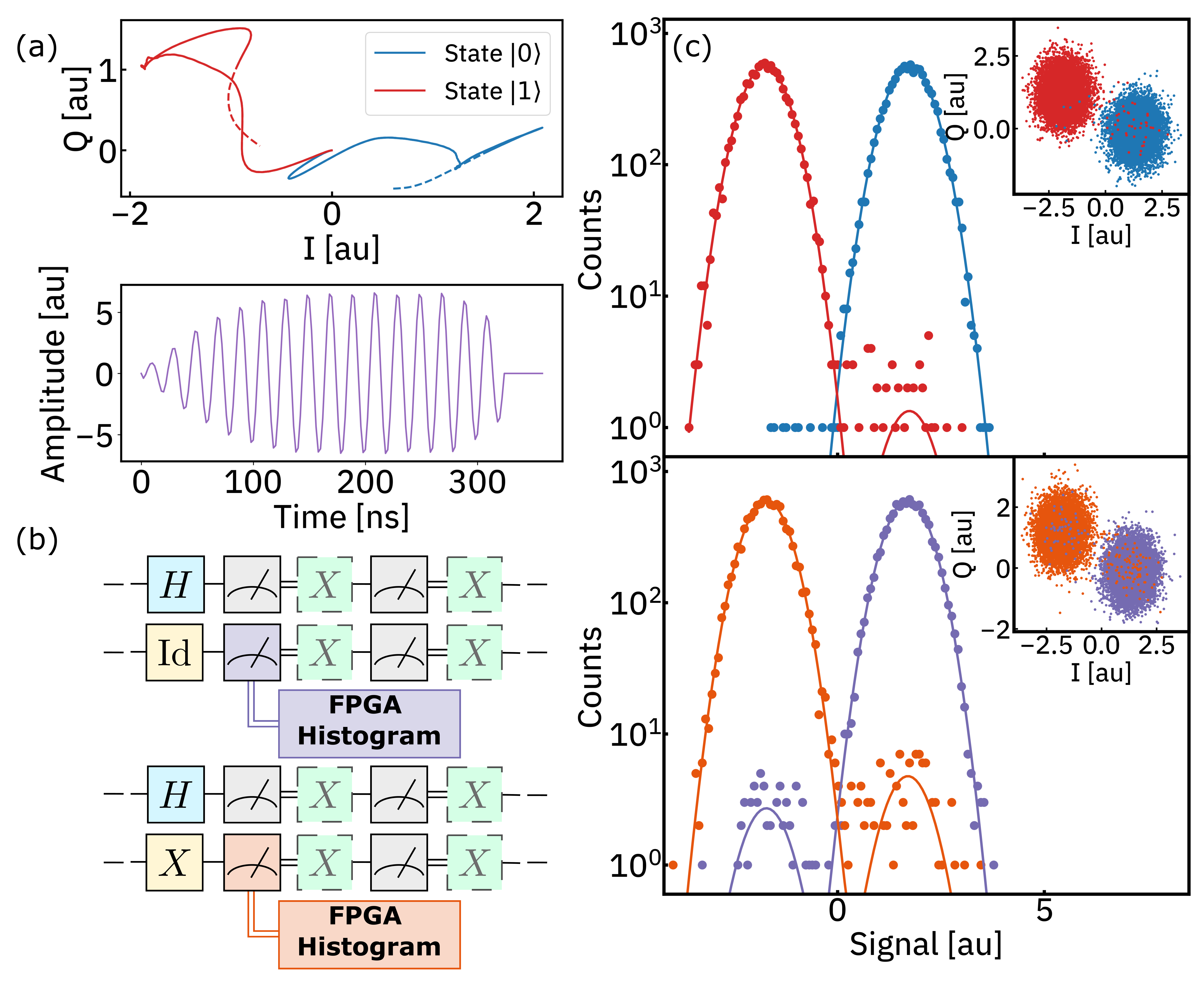}
		\caption{\label{Fig2} Qubit measurement and reset of the pointer qubit. (a) I/Q traces (upper plot) and discriminator kernel (lower plot). The readout signal is sampled for 360 ns (dashed in I/Q traces) and the kernel is applied for 320 ns. (b) Experimental sequences to assess qubit reset quality. The sequences are played in order from top to bottom at a repetition rate of 100 kHz. The Hadamard state preparation sequences are used to prevent error propagation between the $|0\rangle$ and $|1\rangle$ state preparation sequences. The first measurement after each computational state preparation is sampled 10,000 times and used to build the histograms in the bottom panel of (c). The histograms in the top panel of (c) are obtained from alternate preparations of each computational state at 1 kHz experimental repetition rate, allowing the qubit to thermally decay between each experiment. From the histograms we extract an assignment error of 3e-3. The insets show the integrated samples in the I/Q plane. The bottom panel of (c) shows histograms of data taken at 100 kHz repetition rate, with an assignment error of 8.3e-3.}
	\end{figure}
	
	For these experiments, we use two superconducting transmon qubits~\cite{Koch07} in a 14-qubit quantum processor. Both the qubit to qubit coupling and the qubit readouts are mediated by coplanar waveguide resonators~\cite{Blais04}. Qubit readout and conditional reset~\cite{Riste12} are critical parts of this work, and the successful performance of a measurement and reset cycle relies not only on the dynamics of the qubit-resonator interaction, but on efficient classical electronics hardware and software~\cite{Salathe18, Walter17}. Here we use a field programmable gate array (FPGA) platform to measure and control our quantum system, as well as for feedback (here understood as conditional qubit reset following a measurement) and feed-forward (conditional operations as required by the algorithm) \cite{Ryan17}. More details about the electronics used in this work are given in the Supplementary material \cite{Supp}.

		\begin{figure*}[t!]
		\centering
		\includegraphics[width=\textwidth]{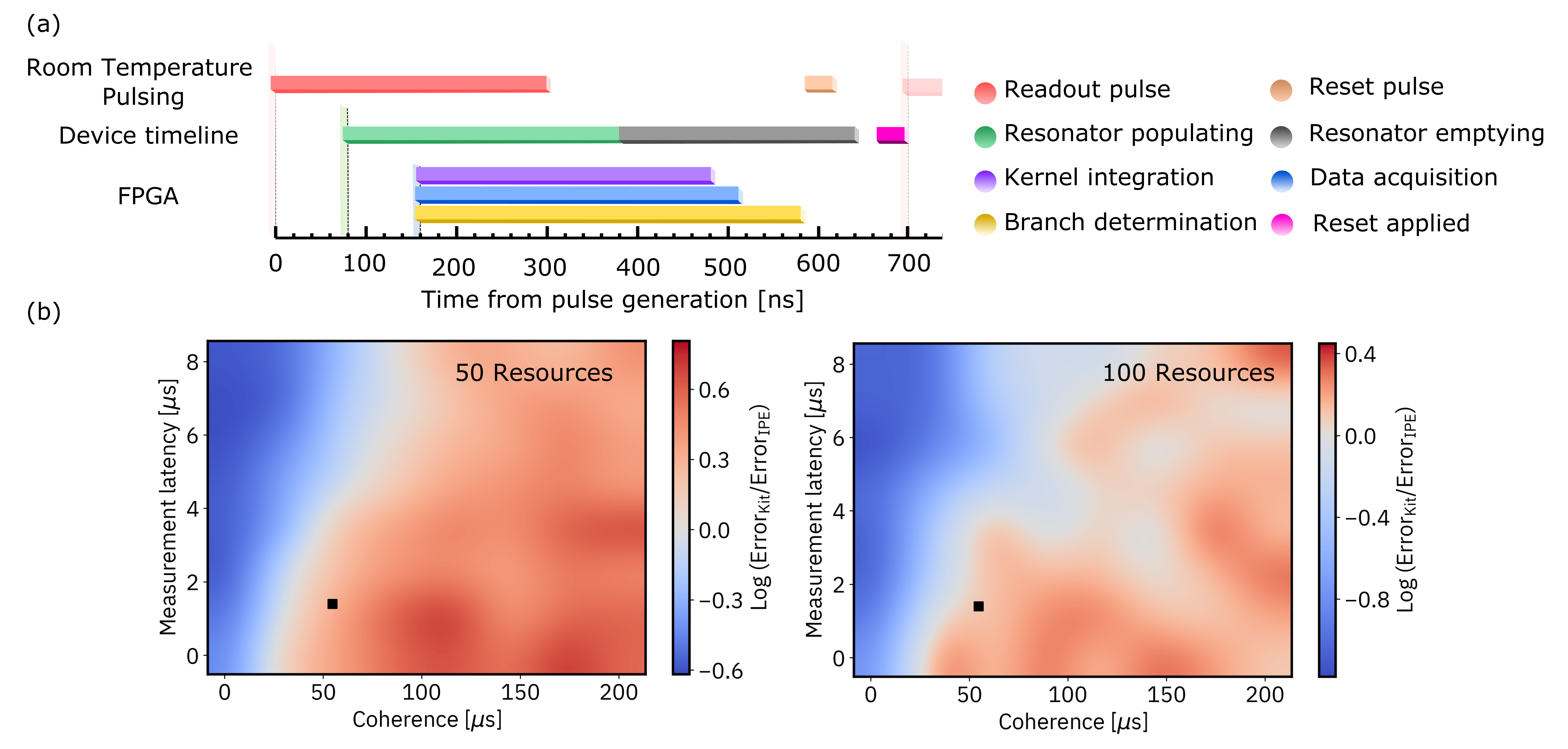}
		\caption{\label{Fig3} (a) Latency breakdown in a measurement and reset cycle. The control electronics entire latency to determine the qubit state, send a conditional marker to a blanking switch, and determine the IPE branching path, is comparable to the latency due to resonator dynamics (population and emptying). This cycle is repeated twice during the algorithm, yielding a total qubit reset latency of 1.4 $\mu$s. (b) Error maps obtained from simulations of both algorithms at difference qubit coherences and reset latencies for 6 bit accuracy and 50 (left) and 100 (right) resources. The black squares correspond to our experimental metrics. In these simulations we use the same qubit gate lengths as in our experiments, and a constant readout and reset error for the pointer qubit of 0.01 and 0.03, respectively.}
	\end{figure*}
	
	The readout and conditional reset of our pointer qubit is a critical aspect of this work. Fig. \ref{Fig2} shows different aspects of our measurement and reset protocol. The returning readout pulse at 7.01325 GHz is down converted to an IF frequency of 50 MHz. We prepare the qubits in the states $|0\rangle$ and $|1\rangle$ and use their readout signal statistics to calibrate a kernel for qubit state discrimination \cite{Ryan15,Magesan15}. Fig. \ref{Fig2}(a, top) shows the I/Q trajectories of the demodulated signals for $|0\rangle$ (blue) and $|1\rangle$ (red). The trajectories are sampled for 360 ns but the optimal kernel is obtained from 4 to 324 ns. Fig. \ref{Fig2}(a, bottom) shows the discriminator kernel, which is computed as the difference between the average signals from the qubit $|0\rangle$ and $|1\rangle$ states. We characterize our readout fidelity by gathering statistics of the readout signal from the two qubit states and binning the integrated shots. The histograms for both states are shown in Fig. \ref{Fig2}(c, top) for $|0\rangle$ (blue) and $|1\rangle$ (red). The inset shows the integrated shots in the I/Q plane. Each state is prepared and measured 10,000 times. In order to minimize state preparation error, these shots are taken at 1 kHz repetition rate, which allows the qubit to thermally relax to its ground state in between experiments. From the histograms we obtain a Fisher separation \cite{Bronn15,Magesan15} of 46 and a readout assignment error of 3e-3.

	We then look at the result of running the same experiments at 100 kHz repetition rate, which is faster than the qubit relaxation time. In this case, we precede each qubit state preparation by an initialization sequence consisting of randomizing the state of the qubit in the standard basis using a Hadamard operation, followed by two conditional qubit resets. The four alternating sequences are shown in Fig. \ref{Fig2}(b), with the corresponding histograms and integrated shots shown in Fig. \ref{Fig2}(c, bottom). From the histograms we obtain $P(0|1)=0.0104$ and $P(1|0)=0.0062$, where $P(i|j)$ is the probability of measuring state $i$ when preparing state $j$. The reset error after two cycles of conditional reset is 0.01 \cite{Supp}.

		\begin{figure*}[t!]
	\centering
	\includegraphics[width=\textwidth]{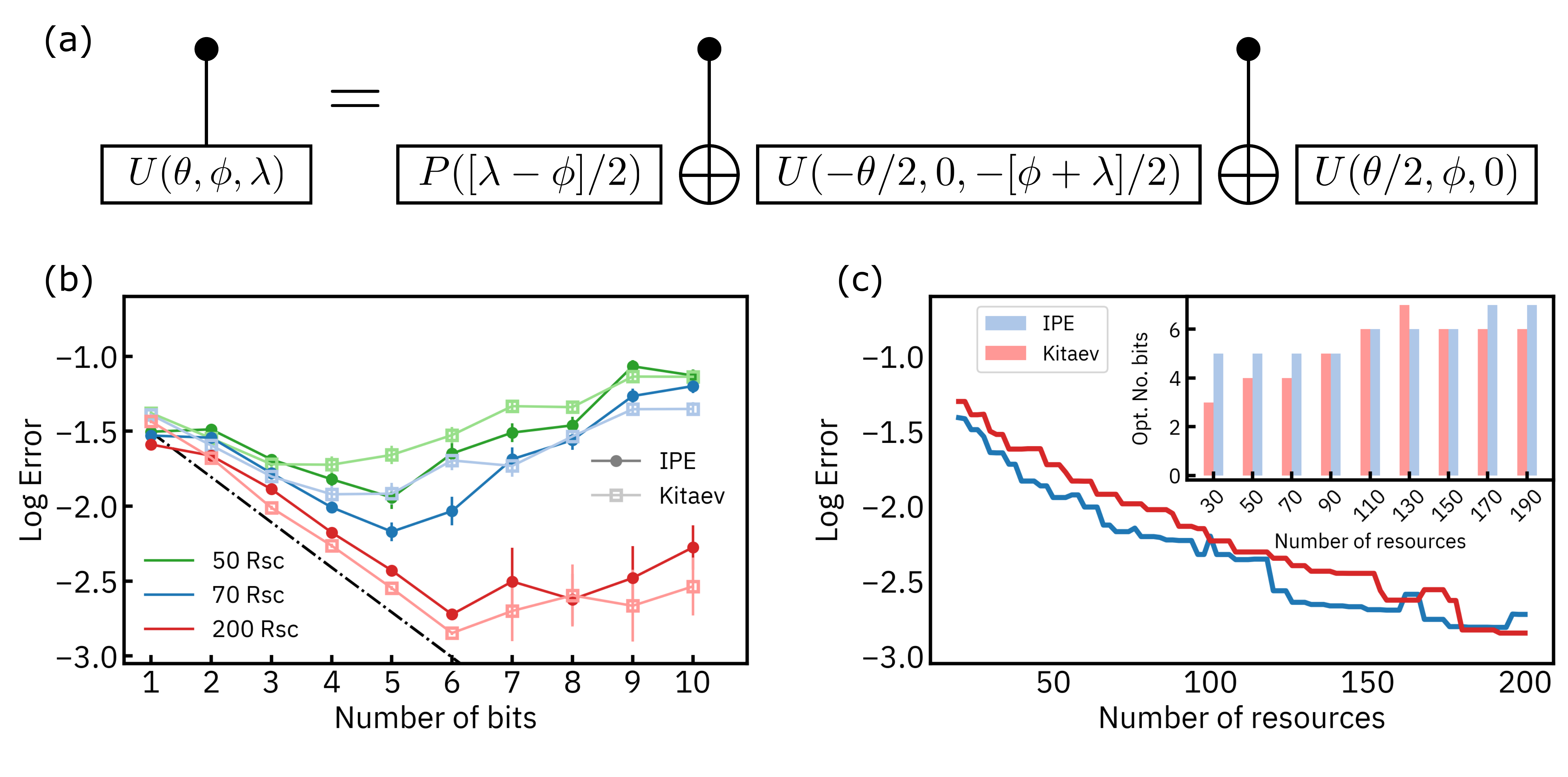}
	\caption{\label{Fig4} (a) Decomposition of a controlled-$U$, where $U$ is a parameterized element of $SU(2)$, in terms of single-qubit unitaries ($U$ rotations and the phase gate $P$) and CNOTs. (b) Phase estimation error up to 10-bit accuracy for both algorithms when the total number of resources are kept constant at 50, 70, and 200. The lower bound of the error is shown for each bit as a dash-dotted black line. Note that due to the nature of the Kitaev estimator, the Kitaev algorithm provides two extra bits of accuracy. (c) Algorithm error as a function of the number of available resources for IPE (blue) and Kitaev (red). The lowest error is chosen among the available number of bits for a given number of resources for both algorithms. The optimal number of bits for a discrete set of resources for both protocols is shown in the inset.}
\end{figure*}
	The real-time computing part of the IPE algorithm has two main components: the application of a bit-flip gate to the pointer qubit conditioned on its measurement yielding the state $|1\rangle$ (qubit reset) and the determination of the $Z-$rotation angle $\theta_k$ for each subsequent circuit. The former determines the measurement and reset latency cycle, which we can define as the elapsed time between the measurement tone being fired at room temperature and the system (room temperature electronics plus quantum processor) being ready for a new operation (gate or measurement). Our experimental latencies for this cycle are shown in Fig. \ref{Fig3}(a). There are two main independent timelines: the readout resonator dynamics and the control electronics latencies and delays. A cable length (plus internal electronics) latency of 160 ns adds an additional constant shift to the relative timings. We have separated the different contributions to the total cycle length vertically in Fig. \ref{Fig3}(a) for clarity. Defining as $t=0$ the instant the 300 ns long readout pulse is sent by the control electronics, we reach the state determination at 488 ns (purple bar). At this time, the FPGA logic operates the blanking switch that screens the reset tone (peach bar), which is applied at 585 ns and reaches the qubit some time later (pink bar). Meanwhile, at the quantum processor, the readout resonator empties of readout photons at a rate determined by the resonator $Q \sim 1150$ (gray bar). Even though the reset pulse is sent at a time when the resonator is still slightly populated, the additional cable and electronics internal latency justifies our placement of the qubit reset pulse at 585 ns. This offers a reasonable trade-off between avoiding both resonator residual population and qubit decay effects. Finally, a few tens of ns are added to the entire cycle due to constrains related to the ADC sampling rate and the need for consistency in the measurement phase for kernel discrimination. It is important that the determination of the conditioned phase rotation happens within the described measurement and reset cycle, so as to not add further delay in the algorithm when implementing this part of the real-time computing. In our experiments, the IPE algorithm is encoded in software as a branching protocol, with the branch determination taking place after each of the IPE circuits shown in Fig. \ref{Fig1} (bottom). We can see in Fig. \ref{Fig3}(a) that the branch determination latency (yellow bar) falls well within the measurement and reset cycle limits.

	For our experiments, we encode these phases as eigenvalues of the Pauli operator $X$. With this choice of $U$ we aim at maximizing the exposure to decoherent noise of the eigenstate-encoding qubit. Once  an operator $U$ for the problem has been chosen, we can implement a controlled-$U$ operation using only CNOT and $SU(2)$ unitaries on the target as shown in Fig. \ref{Fig4}(a). In order to extend our experimental reach and the comparison range for both protocols, we implement each controlled-$U^{2^{k-1}}$ not by repeated applications of controlled-$U$, but by a single application of a controlled-$U^{\prime}$ where $U^{\prime} |u\rangle = e^{2^{k-1}i\phi} |u\rangle$. This is an important requirement, as efficient implementations of powers of U in the QPE algorithm are critical to obtain an exponential speed up.  With this approach, each implementation of the controlled-$U^{2^{k-1}}$ in the circuits in Fig. \ref{Fig1} contains just two CNOTs.

	The key limiting factor we are interested in is the sampling allowance for each circuit in terms of number of allowed resources. We define a resource as the physical act of measuring a qubit. For a given number of available resources $R$ and bits $m$, we sample the circuits in Fig. \ref{Fig1} $\lfloor R/2m \rfloor$ times in the Kitaev case and $\lfloor R/m \rfloor$ times in the IPE case, as Kitaev's protocol calls for 2 measurements per bit. Whereas the Kitaev algorithm relies on sampling statistics, the IPE would require a single measurement in the noiseless case. The presence of noise, however, makes the problem of sampling the IPE circuits slightly less trivial. We have experimentally explored four different ways of determining an IPE solution from its circuit output distributions: taking the expectation value of the entire distribution,  taking the most likely bitstring, and taking the weighted average of the two most likely bitstrings, with and without the constrain that those bitstring are consecutive. We have found that a weighted average to the two most likely \textit{consecutive} bitstrings yielded the lowest errors and therefore use that method for determining the IPE error throughout this work \cite{Supp}.

	We implement the IPE and Kitaev's protocols up to 10 bits and for 600 different phases chosen randomly from the interval $[-\pi, \pi)$. The results for both protocols when limiting the number of resources to 50, 70, and 200 are shown in Fig. \ref{Fig4} (b). We show the error lower bound as a black dash-dotted line. We see, for a given number of resources, that both algorithms show an improvement in the phase estimation accuracy as the number of bits increases. The total resources available keep getting distributed equally among bits and thus above some bit number, the algorithm performance starts to suffer. Besides reducing the number of resources per bit as the number of bits increases, in the case of the IPE algorithm the circuits also become deeper, whereas for Kitaev all the circuits have the same depth independent of the number of bits. This makes IPE to compare unfavorably above certain number of bits. However, if we do not restrict for number of bits and simply consider the algorithm error for a given number of resources, we find that IPE gives consistently lower errors below around 100 resources (Fig. \ref{Fig4} c). In this regime the dynamic circuits exploited by this algorithm offer a clear advantage. As the number of available resources is increased, both algorithms become comparable both in the error and in the optimal number of bits (Fig. \ref{Fig4} (c) inset).
	
	The advantage of exploiting classical electronics within time scales commensurate with qubit lifetimes for the IPE algorithm can be appreciated in Fig. \ref{Fig3} (b). Here we show simulations for both protocols as a function of the quantum system coherence time (where we assume a single metric $T_1 = T_2$ for both qubits, which is not far from typical experimental scenarios) and measurement and reset cycle latency for 6 bits and limiting the number of resources to 50 (left) and 100 (right). The circuits are simulating using the same gate lengths as in our experiments \cite{Supp} and with a constant measurement error of 0.01 and reset error of 0.03. The black square indicates the placement of our experiments, taking the average $T_1$ and $T_2$ for both qubits and 1.4 $\mu$s latency, as we perform each measurement and reset cycle twice for higher reset fidelity. We observe that as the number of resources is allowed to increase, it becomes increasingly demanding to exploit classical electronics for better algorithm performance, although, as shown in Fig. \ref{Fig4} (b), the individual results for both protocols improves.
	
	The experiments shown in this work demonstrate that quantum computing hardware has reached a level of maturity where it can benefit from dynamic circuits. In these circuits the performance of a quantum algorithm can depend on the the classical real-time computing architectures. This demonstration of QPE in a solid-state system can be considered as a first step towards larger scale demonstrations of algorithms that can exploit dynamic circuits and shows that a carefully designed quantum system must take into account all of the components (quantum processor, readout, control electronics, and software). As we build to larger more powerful systems we expect dynamic circuits to be the core for future quantum circuit libraries, algorithms and applications.
	
	We thank Blake Johnson, and Kristan Temme for insightful discussions. We acknowledge partial support from Intelligence Advanced Research Projects Activity (IARPA) under Contract No. W911NF-16-1-0114.

\bibliographystyle{unsrt}
\bibliography{QPE_bib}

\begin{thebibliography}{10}

\bibitem{IQX}
{IBM} {Q}uantum {E}xperience.
\newblock \url{https://quantum-computing.ibm.com}.

\bibitem{Honeywell}
{H}oneywell {Q}uantum {S}olutions.
\newblock \url{https://www.honeywell.com/us/en/company/quantum}.

\bibitem{Allcock11}
D~T~C Allcock, L~Guidoni, T~P Harty, C~J Ballance, M~G Blain, A~M Steane, and
  D~M Lucas.
\newblock Reduction of heating rate in a microfabricated ion trap by
  pulsed-laser cleaning.
\newblock {\em New Journal of Physics}, 13(12):123023, dec 2011.

\bibitem{Gambetta17}
J.~M. {Gambetta}, C.~E. {Murray}, Y.~.~. {Fung}, D.~T. {McClure}, O.~{Dial},
  W.~{Shanks}, J.~W. {Sleight}, and M.~{Steffen}.
\newblock Investigating surface loss effects in superconducting transmon
  qubits.
\newblock {\em IEEE Transactions on Applied Superconductivity}, 27(1):1--5,
  2017.

\bibitem{Wang15}
C.~Wang, C.~Axline, Y.~Y. Gao, T.~Brecht, Y.~Chu, L.~Frunzio, M.~H. Devoret,
  and R.~J. Schoelkopf.
\newblock Surface participation and dielectric loss in superconducting qubits.
\newblock {\em Applied Physics Letters}, 107(16):162601, 2015.

\bibitem{Harty14}
T.~P. Harty, D.~T.~C. Allcock, C.~J. Ballance, L.~Guidoni, H.~A. Janacek, N.~M.
  Linke, D.~N. Stacey, and D.~M. Lucas.
\newblock High-fidelity preparation, gates, memory, and readout of a
  trapped-ion quantum bit.
\newblock {\em Phys. Rev. Lett.}, 113:220501, Nov 2014.

\bibitem{Ospelkaus11}
C.~Ospelkaus, U.~Warring, Y.~Colombe, K.~R. Brown, J.~M. Amini, D.~Leibfried,
  and D.~J. Wineland.
\newblock Microwave quantum logic gates for trapped ions.
\newblock {\em Nature}, 476(7359):181--184, 2011.

\bibitem{Chow09}
J.~M. Chow, J.~M. Gambetta, L.~Tornberg, Jens Koch, Lev~S. Bishop, A.~A. Houck,
  B.~R. Johnson, L.~Frunzio, S.~M. Girvin, and R.~J. Schoelkopf.
\newblock Randomized benchmarking and process tomography for gate errors in a
  solid-state qubit.
\newblock {\em Phys. Rev. Lett.}, 102:090502, Mar 2009.

\bibitem{Lucero10}
Erik Lucero, Julian Kelly, Radoslaw~C. Bialczak, Mike Lenander, Matteo
  Mariantoni, Matthew Neeley, A.~D. O'Connell, Daniel Sank, H.~Wang, Martin
  Weides, James Wenner, Tsuyoshi Yamamoto, A.~N. Cleland, and John~M. Martinis.
\newblock Reduced phase error through optimized control of a superconducting
  qubit.
\newblock {\em Phys. Rev. A}, 82:042339, Oct 2010.

\bibitem{Riste12}
D.~Rist\`e, C.~C. Bultink, K.~W. Lehnert, and L.~DiCarlo.
\newblock Feedback control of a solid-state qubit using high-fidelity
  projective measurement.
\newblock {\em Phys. Rev. Lett.}, 109:240502, Dec 2012.

\bibitem{Walter17}
T.~Walter, P.~Kurpiers, S.~Gasparinetti, P.~Magnard,
  A.~Poto\ifmmode~\check{c}\else \v{c}\fi{}nik, Y.~Salath\'e, M.~Pechal,
  M.~Mondal, M.~Oppliger, C.~Eichler, and A.~Wallraff.
\newblock Rapid high-fidelity single-shot dispersive readout of superconducting
  qubits.
\newblock {\em Phys. Rev. Applied}, 7:054020, May 2017.

\bibitem{Chow11}
Jerry~M. Chow, A.~D. C\'orcoles, Jay~M. Gambetta, Chad Rigetti, B.~R. Johnson,
  John~A. Smolin, J.~R. Rozen, George~A. Keefe, Mary~B. Rothwell, Mark~B.
  Ketchen, and M.~Steffen.
\newblock Simple all-microwave entangling gate for fixed-frequency
  superconducting qubits.
\newblock {\em Phys. Rev. Lett.}, 107:080502, Aug 2011.

\bibitem{Barends14}
R.~Barends, J.~Kelly, A.~Megrant, A.~Veitia, D.~Sank, E.~Jeffrey, T.~C. White,
  J.~Mutus, A.~G. Fowler, B.~Campbell, Y.~Chen, Z.~Chen, B.~Chiaro,
  A.~Dunsworth, C.~Neill, P.~O'Malley, P.~Roushan, A.~Vainsencher, J.~Wenner,
  A.~N. Korotkov, A.~N. Cleland, and John~M. Martinis.
\newblock Superconducting quantum circuits at the surface code threshold for
  fault tolerance.
\newblock {\em Nature}, 508(7497):500--503, 2014.

\bibitem{Ballance16}
C.~J. Ballance, T.~P. Harty, N.~M. Linke, M.~A. Sepiol, and D.~M. Lucas.
\newblock High-fidelity quantum logic gates using trapped-ion hyperfine qubits.
\newblock {\em Phys. Rev. Lett.}, 117:060504, Aug 2016.

\bibitem{Gambetta12}
Jay~M. Gambetta, A.~D. C\'orcoles, S.~T. Merkel, B.~R. Johnson, John~A. Smolin,
  Jerry~M. Chow, Colm~A. Ryan, Chad Rigetti, S.~Poletto, Thomas~A. Ohki,
  Mark~B. Ketchen, and M.~Steffen.
\newblock Characterization of addressability by simultaneous randomized
  benchmarking.
\newblock {\em Phys. Rev. Lett.}, 109:240504, Dec 2012.

\bibitem{Takita16}
Maika Takita, A.~D. C\'orcoles, Easwar Magesan, Baleegh Abdo, Markus Brink,
  Andrew Cross, Jerry~M. Chow, and Jay~M. Gambetta.
\newblock Demonstration of weight-four parity measurements in the surface code
  architecture.
\newblock {\em Phys. Rev. Lett.}, 117:210505, Nov 2016.

\bibitem{McKay20}
David~C. McKay, Andrew~W. Cross, Christopher~J. Wood, and Jay~M. Gambetta.
\newblock Correlated randomized benchmarking, 2020.

\bibitem{havlicek19}
Vojt{\v e}ch Havl{\'\i}{\v c}ek, Antonio~D. C{\'o}rcoles, Kristan Temme,
  Aram~W. Harrow, Abhinav Kandala, Jerry~M. Chow, and Jay~M. Gambetta.
\newblock Supervised learning with quantum-enhanced feature spaces.
\newblock {\em Nature}, 567(7747):209--212, 2019.

\bibitem{Kandala17}
Abhinav Kandala, Antonio Mezzacapo, Kristan Temme, Maika Takita, Markus Brink,
  Jerry~M. Chow, and Jay~M. Gambetta.
\newblock Hardware-efficient variational quantum eigensolver for small
  molecules and quantum magnets.
\newblock {\em Nature}, 549(7671):242--246, 2017.

\bibitem{Cross19}
Andrew~W. Cross, Lev~S. Bishop, Sarah Sheldon, Paul~D. Nation, and Jay~M.
  Gambetta.
\newblock Validating quantum computers using randomized model circuits.
\newblock {\em Phys. Rev. A}, 100:032328, Sep 2019.

\bibitem{Griffiths96}
Robert~B. Griffiths and Chi-Sheng Niu.
\newblock Semiclassical fourier transform for quantum computation.
\newblock {\em Phys. Rev. Lett.}, 76:3228--3231, Apr 1996.

\bibitem{Minev19}
Z.K. Minev, S.O. Mundhada, S.~Shankar, P.~Reinhold, R.~Gutierrez-Jauregui, R.J.
  Schoelkopf, M.~Mirrahimi, H.J. Carmichael, and M.H. Devoret.
\newblock To catch and reverse a quantum jump mid-flight.
\newblock {\em Nature}, (570):200--204, 2019.

\bibitem{Ofek16}
Nissim Ofek, Andrei Petrenko, Reinier Heeres, Philip Reinhold, Zaki Leghtas,
  Brian Vlastakis, Yehan Liu, Luigi Frunzio, S.~M. Girvin, L.~Jiang, Mazyar
  Mirrahimi, M.~H. Devoret, and R.~J. Schoelkopf.
\newblock Extending the lifetime of a quantum bit with error correction in
  superconducting circuits.
\newblock {\em Nature}, 536(7617):441--445, 2016.

\bibitem{Andersen19}
Christian~Kraglund Andersen, Ants Remm, Stefania Lazar, Sebastian Krinner,
  Johannes Heinsoo, Jean-Claude Besse, Mihai Gabureac, Andreas Wallraff, and
  Christopher Eichler.
\newblock Entanglement stabilization using ancilla-based parity detection and
  real-time feedback in superconducting circuits.
\newblock {\em npj Quantum Information}, 5(1):69, 2019.

\bibitem{Barrett04}
M.~D. Barrett, J.~Chiaverini, T.~Schaetz, J.~Britton, W.~M. Itano, J.~D. Jost,
  E.~Knill, C.~Langer, D.~Leibfried, R.~Ozeri, and D.~J. Wineland.
\newblock Deterministic quantum teleportation of atomic qubits.
\newblock {\em Nature}, 429(6993):737--739, 2004.

\bibitem{Riebe04}
M.~Riebe, H.~H{\"a}ffner, C.~F. Roos, W.~H{\"a}nsel, J.~Benhelm, G.~P.~T.
  Lancaster, T.~W. K{\"o}rber, C.~Becher, F.~Schmidt-Kaler, D.~F.~V. James, and
  R.~Blatt.
\newblock Deterministic quantum teleportation with atoms.
\newblock {\em Nature}, 429(6993):734--737, 2004.

\bibitem{Steffen13}
L.~Steffen, Y.~Salathe, M.~Oppliger, P.~Kurpiers, M.~Baur, C.~Lang, C.~Eichler,
  G.~Puebla-Hellmann, A.~Fedorov, and A.~Wallraff.
\newblock Deterministic quantum teleportation with feed-forward in a solid
  state system.
\newblock {\em Nature}, 500(7462):319--322, 2013.

\bibitem{Chou18}
Kevin~S. Chou, Jacob~Z. Blumoff, Christopher~S. Wang, Philip~C. Reinhold,
  Christopher~J. Axline, Yvonne~Y. Gao, L.~Frunzio, M.~H. Devoret, Liang Jiang,
  and R.~J. Schoelkopf.
\newblock Deterministic teleportation of a quantum gate between two logical
  qubits.
\newblock {\em Nature}, 561(7723):368--373, 2018.

\bibitem{Ryan17}
Colm~A. Ryan, Blake~R. Johnson, Diego Ristè, Brian Donovan, and Thomas~A.
  Ohki.
\newblock Hardware for dynamic quantum computing.
\newblock {\em Review of Scientific Instruments}, 88(10):104703, 2017.

\bibitem{Reinhold20}
Philip Reinhold, Serge Rosenblum, Wen-Long Ma, Luigi Frunzio, Liang Jiang, and
  Robert~J. Schoelkopf.
\newblock Error-corrected gates on an encoded qubit.
\newblock {\em Nature Physics}, 16(8):822--826, 2020.

\bibitem{nielsen00}
Michael~A. Nielsen and Isaac~L. Chuang.
\newblock {\em Quantum Computation and Quantum Information}.
\newblock Cambridge University Press, 2000.

\bibitem{Cleve98}
R.~Cleve, A.~Ekert, C.~Macchiavello, and M.~Mosca.
\newblock Quantum algorithms revisited.
\newblock {\em Proc. R. Soc. Lond. A}, 454:339--354, 1998.

\bibitem{Barenco96}
Adriano Barenco, Artur Ekert, Kalle-Antti Suominen, and P\"aivi T\"orm\"a.
\newblock Approximate quantum fourier transform and decoherence.
\newblock {\em Phys. Rev. A}, 54:139--146, Jul 1996.

\bibitem{Kitaev95}
A.~Yu. Kitaev.
\newblock Quantum measurements and the abelian stabilizer problem, 1995.

\bibitem{Childs2000}
Andrew~M. Childs, John Preskill, and Joseph Renes.
\newblock Quantum information and precision measurement.
\newblock {\em Journal of Modern Optics}, 47(2-3):155--176, 2000.

\bibitem{Knill07}
Emanuel Knill, Gerardo Ortiz, and Rolando~D. Somma.
\newblock Optimal quantum measurements of expectation values of observables.
\newblock {\em Phys. Rev. A}, 75:012328, Jan 2007.

\bibitem{Dobsicek07}
Miroslav Dob\ifmmode \check{s}\else \v{s}\fi{}\'{\i}\ifmmode~\check{c}\else
  \v{c}\fi{}ek, G\"oran Johansson, Vitaly Shumeiko, and G\"oran Wendin.
\newblock Arbitrary accuracy iterative quantum phase estimation algorithm using
  a single ancillary qubit: A two-qubit benchmark.
\newblock {\em Phys. Rev. A}, 76:030306, Sep 2007.

\bibitem{Higgins09}
B~L Higgins, D~W Berry, S~D Bartlett, M~W Mitchell, H~M Wiseman, and G~J Pryde.
\newblock Demonstrating heisenberg-limited unambiguous phase estimation without
  adaptive measurements.
\newblock {\em New Journal of Physics}, 11(7):073023, jul 2009.

\bibitem{Wiebe16}
Nathan Wiebe and Chris Granade.
\newblock Efficient bayesian phase estimation.
\newblock {\em Phys. Rev. Lett.}, 117:010503, Jun 2016.

\bibitem{Paesani17}
S.~Paesani, A.~A. Gentile, R.~Santagati, J.~Wang, N.~Wiebe, D.~P. Tew, J.~L.
  O'Brien, and M.~G. Thompson.
\newblock Experimental bayesian quantum phase estimation on a silicon photonic
  chip.
\newblock {\em Phys. Rev. Lett.}, 118:100503, Mar 2017.

\bibitem{O_Brien_2019}
Thomas~E O'Brien, Brian Tarasinski, and Barbara~M Terhal.
\newblock Quantum phase estimation of multiple eigenvalues for small-scale
  (noisy) experiments.
\newblock {\em New Journal of Physics}, 21(2):023022, Feb 2019.

\bibitem{Svore14}
Krysta~M. Svore, Matthew~B. Hastings, and Michael Freedman.
\newblock Faster phase estimation.
\newblock {\em Quantum Info. Comput.}, 14(3--4):306--328, March 2014.

\bibitem{Supp}
{S}ee supplementary material.

\bibitem{Kaye07}
Phillip Kaye, Raymond Laflamme, and Michele Mosca.
\newblock {\em An Introduction to Quantum Computing}.
\newblock Oxford University Press, Inc., USA, 2007.

\bibitem{Koch07}
Jens Koch, Terri~M. Yu, Jay Gambetta, A.~A. Houck, D.~I. Schuster, J.~Majer,
  Alexandre Blais, M.~H. Devoret, S.~M. Girvin, and R.~J. Schoelkopf.
\newblock Charge-insensitive qubit design derived from the cooper pair box.
\newblock {\em Phys. Rev. A}, 76:042319, Oct 2007.

\bibitem{Blais04}
Alexandre Blais, Ren-Shou Huang, Andreas Wallraff, S.~M. Girvin, and R.~J.
  Schoelkopf.
\newblock Cavity quantum electrodynamics for superconducting electrical
  circuits: An architecture for quantum computation.
\newblock {\em Phys. Rev. A}, 69:062320, Jun 2004.

\bibitem{Salathe18}
Yves Salath\'e, Philipp Kurpiers, Thomas Karg, Christian Lang,
  Christian~Kraglund Andersen, Abdulkadir Akin, Sebastian Krinner, Christopher
  Eichler, and Andreas Wallraff.
\newblock Low-latency digital signal processing for feedback and feedforward in
  quantum computing and communication.
\newblock {\em Phys. Rev. Applied}, 9:034011, Mar 2018.

\bibitem{Ryan15}
Colm~A. Ryan, Blake~R. Johnson, Jay~M. Gambetta, Jerry~M. Chow, Marcus~P.
  da~Silva, Oliver~E. Dial, and Thomas~A. Ohki.
\newblock Tomography via correlation of noisy measurement records.
\newblock {\em Phys. Rev. A}, 91:022118, Feb 2015.

\bibitem{Magesan15}
Easwar Magesan, Jay~M. Gambetta, A.~D. C\'orcoles, and Jerry~M. Chow.
\newblock Machine learning for discriminating quantum measurement trajectories
  and improving readout.
\newblock {\em Phys. Rev. Lett.}, 114:200501, May 2015.

\bibitem{Bronn15}
N.~T. {Bronn}, E.~{Magesan}, N.~A. {Masluk}, J.~M. {Chow}, J.~M. {Gambetta},
  and M.~{Steffen}.
\newblock Reducing spontaneous emission in circuit quantum electrodynamics by a
  combined readout/filter technique.
\newblock {\em IEEE Transactions on Applied Superconductivity}, 25(5):1--10,
  2015.

\end{thebibliography}


\begin{thebibliography}{10}

\bibitem{Koch07}
Jens Koch, Terri~M. Yu, Jay Gambetta, A.~A. Houck, D.~I. Schuster, J.~Majer,
  Alexandre Blais, M.~H. Devoret, S.~M. Girvin, and R.~J. Schoelkopf.
\newblock Charge-insensitive qubit design derived from the cooper pair box.
\newblock {\em Phys. Rev. A}, 76:042319, Oct 2007.

\bibitem{Blais04}
Alexandre Blais, Ren-Shou Huang, Andreas Wallraff, S.~M. Girvin, and R.~J.
  Schoelkopf.
\newblock Cavity quantum electrodynamics for superconducting electrical
  circuits: An architecture for quantum computation.
\newblock {\em Phys. Rev. A}, 69:062320, Jun 2004.

\bibitem{Klimov18}
P.~V. Klimov, J.~Kelly, Z.~Chen, M.~Neeley, A.~Megrant, B.~Burkett, R.~Barends,
  K.~Arya, B.~Chiaro, Yu~Chen, A.~Dunsworth, A.~Fowler, B.~Foxen, C.~Gidney,
  M.~Giustina, R.~Graff, T.~Huang, E.~Jeffrey, Erik Lucero, J.~Y. Mutus,
  O.~Naaman, C.~Neill, C.~Quintana, P.~Roushan, Daniel Sank, A.~Vainsencher,
  J.~Wenner, T.~C. White, S.~Boixo, R.~Babbush, V.~N. Smelyanskiy, H.~Neven,
  and John~M. Martinis.
\newblock Fluctuations of energy-relaxation times in superconducting qubits.
\newblock {\em Phys. Rev. Lett.}, 121:090502, Aug 2018.

\bibitem{Motzoi09}
F.~Motzoi, J.~M. Gambetta, P.~Rebentrost, and F.~K. Wilhelm.
\newblock Simple pulses for elimination of leakage in weakly nonlinear qubits.
\newblock {\em Phys. Rev. Lett.}, 103:110501, Sep 2009.

\bibitem{McKay17}
David~C. McKay, Christopher~J. Wood, Sarah Sheldon, Jerry~M. Chow, and Jay~M.
  Gambetta.
\newblock Efficient $z$ gates for quantum computing.
\newblock {\em Phys. Rev. A}, 96:022330, Aug 2017.

\bibitem{Chow11}
Jerry~M. Chow, A.~D. C\'orcoles, Jay~M. Gambetta, Chad Rigetti, B.~R. Johnson,
  John~A. Smolin, J.~R. Rozen, George~A. Keefe, Mary~B. Rothwell, Mark~B.
  Ketchen, and M.~Steffen.
\newblock Simple all-microwave entangling gate for fixed-frequency
  superconducting qubits.
\newblock {\em Phys. Rev. Lett.}, 107:080502, Aug 2011.

\bibitem{Sheldon16}
Sarah Sheldon, Easwar Magesan, Jerry~M. Chow, and Jay~M. Gambetta.
\newblock Procedure for systematically tuning up cross-talk in the
  cross-resonance gate.
\newblock {\em Phys. Rev. A}, 93:060302, Jun 2016.

\bibitem{Sundaresan20}
Neereja Sundaresan, Isaac Lauer, Emily Pritchett, Easwar Magesan, Petar
  Jurcevic, and Jay~M. Gambetta.
\newblock Reducing unitary and spectator errors in cross resonance with
  optimized rotary echoes.

\bibitem{Magesan11}
Easwar Magesan, J.~M. Gambetta, and Joseph Emerson.
\newblock Scalable and robust randomized benchmarking of quantum processes.
\newblock {\em Phys. Rev. Lett.}, 106:180504, May 2011.

\bibitem{Corcoles13}
A.~D. C\'orcoles, Jay~M. Gambetta, Jerry~M. Chow, John~A. Smolin, Matthew Ware,
  Joel Strand, B.~L.~T. Plourde, and M.~Steffen.
\newblock Process verification of two-qubit quantum gates by randomized
  benchmarking.
\newblock {\em Phys. Rev. A}, 87:030301, Mar 2013.

\bibitem{McKay19}
David~C. McKay, Sarah Sheldon, John~A. Smolin, Jerry~M. Chow, and Jay~M.
  Gambetta.
\newblock Three-qubit randomized benchmarking.
\newblock {\em Phys. Rev. Lett.}, 122:200502, May 2019.

\bibitem{Yaakobi13}
O.~Yaakobi, L.~Friedland, C.~Macklin, and I.~Siddiqi.
\newblock Parametric amplification in josephson junction embedded transmission
  lines.
\newblock {\em Phys. Rev. B}, 87:144301, Apr 2013.

\bibitem{Schuster2005-ACStark}
D~I Schuster, A~Wallraff, A~Blais, L~Frunzio, R.-S. Huang, J~Majer, S~M Girvin,
  and R~J Schoelkopf.
\newblock ac stark shift and dephasing of a superconducting qubit strongly
  coupled to a cavity field.
\newblock {\em Physical Review Letters}, 94(12):123602, Mar 2005.

\bibitem{Gambetta2006-dephasing}
Jay Gambetta, Alexandre Blais, D~I Schuster, A~Wallraff, L~Frunzio, J~Majer,
  M~H Devoret, S~M Girvin, and R~J Schoelkopf.
\newblock Qubit-photon interactions in a cavity: Measurement-induced dephasing
  and number splitting.
\newblock {\em Physical Review A}, 74(4):042318, Oct 2006.

\bibitem{Boissonneault2008}
Maxime Boissonneault, J~M Gambetta, and Alexandre Blais.
\newblock Nonlinear dispersive regime of cavity qed: The dressed dephasing
  model.
\newblock {\em Physical Review A}, 77(6):060305, Jun 2008.

\bibitem{Gambetta2008-qm-traj}
Jay Gambetta, Alexandre Blais, M~Boissonneault, A~A Houck, D~I Schuster, and
  S~M Girvin.
\newblock Quantum trajectory approach to circuit qed: Quantum jumps and the
  zeno effect.
\newblock {\em Physical Review A}, 77(1):012112, Jan 2008.

\bibitem{Boissonneault2009-Photon-induced-relax}
Maxime Boissonneault, J~M Gambetta, and Alexandre Blais.
\newblock Dispersive regime of circuit qed: Photon-dependent qubit dephasing
  and relaxation rates.
\newblock {\em Physical Review A}, 79(1):013819, Jan 2009.

\bibitem{McClure16}
D.~T. McClure, Hanhee Paik, L.~S. Bishop, M.~Steffen, Jerry~M. Chow, and Jay~M.
  Gambetta.
\newblock Rapid driven reset of a qubit readout resonator.
\newblock {\em Phys. Rev. Applied}, 5:011001, Jan 2016.

\bibitem{Macklin15}
C.~Macklin, K.~O{\textquoteright}Brien, D.~Hover, M.~E. Schwartz,
  V.~Bolkhovsky, X.~Zhang, W.~D. Oliver, and I.~Siddiqi.
\newblock A near{\textendash}quantum-limited josephson traveling-wave
  parametric amplifier.
\newblock {\em Science}, 350(6258):307--310, 2015.

\bibitem{Bronn17}
N~T Bronn, B~Abdo, K~Inoue, S~Lekuch, A~D C{\'{o}}rcoles, J~B Hertzberg,
  M~Takita, L~S Bishop, J~M Gambetta, and J~M Chow.
\newblock Fast, high-fidelity readout of multiple qubits.
\newblock {\em Journal of Physics: Conference Series}, 834:012003, may 2017.

\bibitem{Ryan:2015}
Colm~A. Ryan, Blake~R. Johnson, Jay~M. Gambetta, Jerry~M. Chow, Marcus~P.
  da~Silva, Oliver~E. Dial, and Thomas~A. Ohki.
\newblock Tomography via correlation of noisy measurement records.
\newblock {\em Phys. Rev. A}, 91:022118, Feb 2015.

\bibitem{nielsen00}
Michael~A. Nielsen and Isaac~L. Chuang.
\newblock {\em Quantum Computation and Quantum Information}.
\newblock Cambridge University Press, 2000.

\bibitem{Cleve98}
R.~Cleve, A.~Ekert, C.~Macchiavello, and M.~Mosca.
\newblock Quantum algorithms revisited.
\newblock {\em Proc. R. Soc. Lond. A}, 454:339--354, 1998.

\bibitem{Kaye07}
Phillip Kaye, Raymond Laflamme, and Michele Mosca.
\newblock {\em An Introduction to Quantum Computing}.
\newblock Oxford University Press, Inc., USA, 2007.

\end{thebibliography}
	
\end{document}


\onecolumngrid
	
	\section*{ Supplementary Information: \\
		\vspace{0.3cm} Exploiting dynamic quantum circuits in a quantum algorithm with superconducting qubits}
	
	\beginsupplement
	\subsection*{Device characterization}
	Our device consists of 14 superconducting transmon qubits \cite{Koch07} coupled by coplanar waveguide (CPW) resonators and read out via the cQED architecture \cite{Blais04}. The qubits are laid out in a topology that contains both degree-2 and degree-3 connectivities, meaning each qubit has either two or three nearest neighbors. The two qubits used in this work, $Q_0$ (the `system' qubit) and $Q_1$ (the `pointer' qubit), have two nearest neighbors each, including each other. The fundamental frequency of $Q_0$ ($Q_1$) is $\omega_{\mathrm{ge}}/2\pi = 5.1566$ (5.3634) GHz, with anharmonicity $\alpha = \omega_{\mathrm{ef}}/2\pi - \omega_{\mathrm{ge}}/2\pi = -341.8$ (-343.1) MHz and the qubit is read out at a frequency of 7.099091 (7.013250) GHz. The relaxation and dephasing experiments (around 150 per metric and qubit) used to extract these values were interleaved in between the main phase estimation experiments discussed in this work. The $T_1$ median value is 68.20 (49.23) $\mu$s for $Q_0$ ($Q_1$), with standard deviation 15.19 (7.75) $\mu$s, whereas for $T_2^{\mathrm{echo}}$, the median value is 59.93 (41.92) $\mu$s with standard deviation 12.71 (8.05) $\mu$s.\\
	\begin{figure}[H]
		\begin{centering}
			\includegraphics[width=\textwidth]{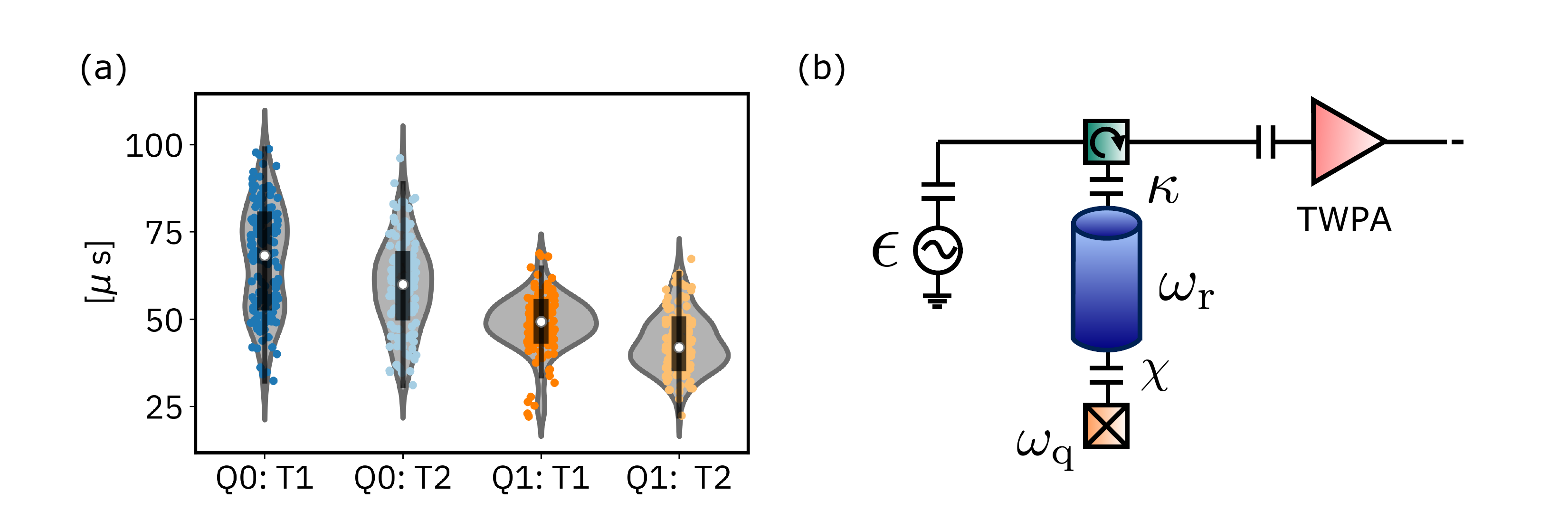}
			\par\end{centering}
		\caption{(a) Lifetime and coherence statistics for both qubits. Each data point was obtained from experiments interleaved between the phase estimation main experiments, to capture in as much as possible, any drift during that time window. (b) Schematic of the measurement apparatus in our experiments. The readout resonator is probed in reflection. A Traveling Wave Parametric Amplifier (TWPA) is used for quantum-limited amplification. \label{coherences}}
		
	\end{figure}
	We show statistics of qubit relaxation time $T_1$ and coherence time $T_2^{\mathrm{echo}}$ in Fig. \ref{coherences}. The distributions sometimes exhibit non-Gaussian shapes (most remarkably $T_1$ for $Q_0$, which follows a double peak distribution). This large variation in qubit lifetimes has been repeatedly observed in superconducting qubits \cite{Klimov18}. \\
	Single-qubit gates are Gaussian-shaped pulses of length 30 ns and $\sigma=7.5$ ns with a 10 ns idle time buffer at the end of each gate to prevent errors arising from evanescent waves in the different cryogenic components present along the drive coaxial lines. These single-qubit gates are implemented using DRAG pulses \cite{Motzoi09} to address errors arising from higher excitation levels in the transmon. Phase rotations are implemented using frame changes in software \cite{McKay17} and are both essentially instantaneous and free of errors. Two-qubit gates are implemented using cross-resonance \cite{Chow11, Sheldon16}, with $Q_1$ (pointer qubit) acting as the control qubit and $Q_0$ (system qubit) as the target qubit, which is the natural way the quantum phase estimation protocols discussed in this work call for the CNOT operations. The cross-resonance tone is implemented using an echo sequence with rotary pulses on the target qubit \cite{Sundaresan20}. Each physical cross-resonance tone, of length 90 ns followed by a 10 ns idle time buffer, is a square-topped Gaussian pulse with 30 ns risetime and falltime. This cross-resonance sequence implements a CNOT up to one control and target pre-rotation, making the total CNOT length 280 ns.\\
	We run simultaneous single-qubit randomized benchmarking \cite{Magesan11} using the basis $\{I, X(+\pi/2), X(-\pi/2),Y(+\pi/2),Y(-\pi/2),Z(+\pi/2),Z(-\pi/2),Z(+\pi)\}$, where $I$ is the identity operation and $R(\theta)$ represents a rotation of angle $\theta$ around axis $R$. With this basis, one can implement each of the 24 single-qubit Clifford operations with an average of one physical gate (ie, $X-$ or $Y-$rotations) per Clifford. For two-qubit randomized benchmarking \cite{Corcoles13} we use the aforementioned single-qubit rotations plus a CNOT gate. This yields an average of 1.5 CNOT gates per two-qubit Clifford. We obtain $7.59\times 10^{-4}\pm1.4\times 10^{-5}$ and $5.67\times 10^{-4}\pm8.1\times 10^{-6}$ error per gate for $Q_0$ and $Q_1$, respectively, over 183 different random sequences running from 1 to 6000 Cliffords, and using a simultaneous single-qubit calibration routine for both qubits \cite{McKay19}. For two-qubits, we obtain $2.32\times 10^{-2}\pm2.0\times 10^{-4}$ error per Clifford. It is a fair assumption in this case to consider that most of that error comes from the CNOT gate. At a weight of 1.5 CNOT per Clifford, we can estimate and approximate CNOT error of $1.55 \times 10^{-2}$.
	
	\begin{figure}[H]
		\begin{centering}
			\includegraphics[width=\textwidth]{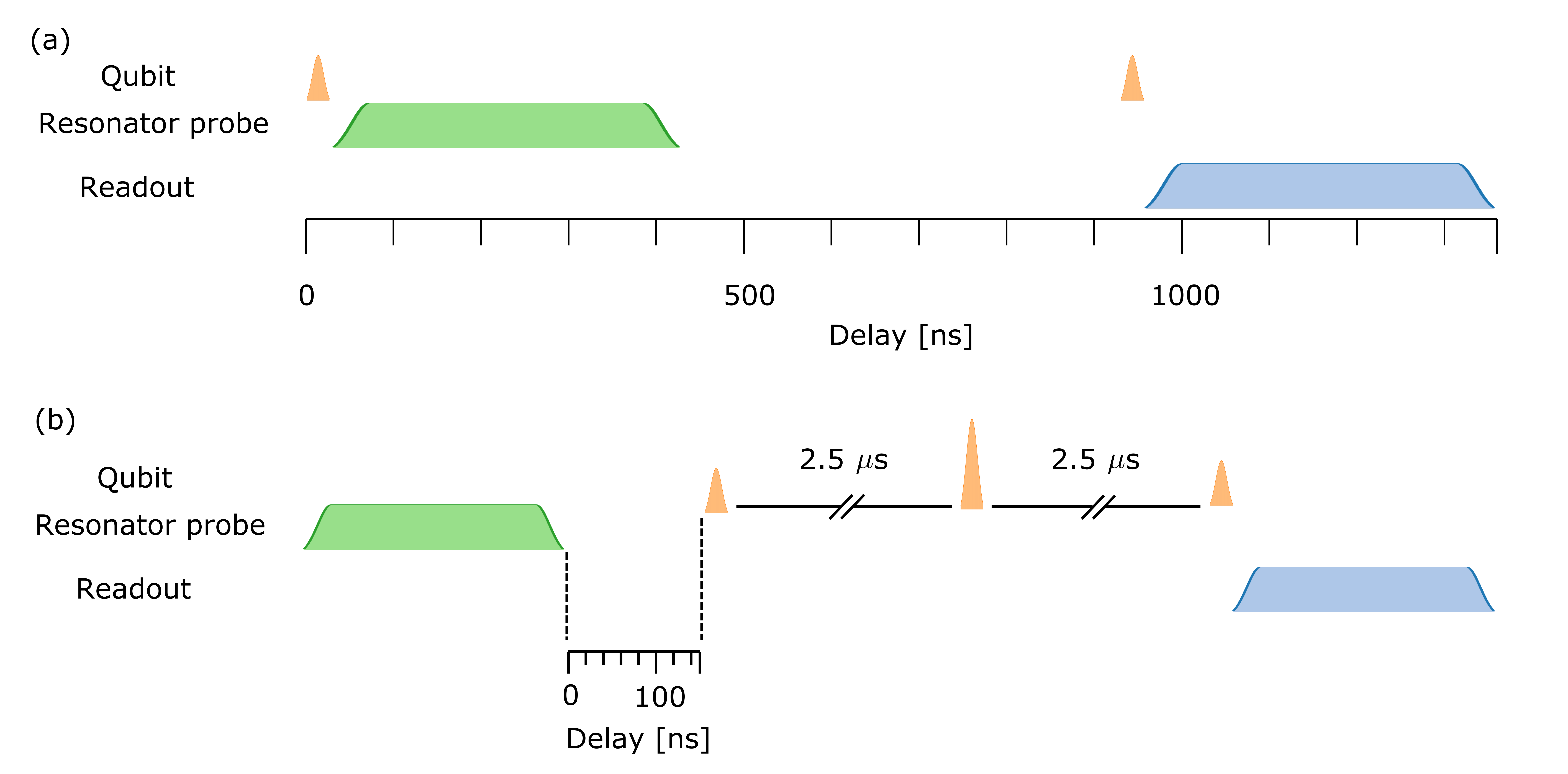}
			\par\end{centering}
		\caption{Schematic for the dressed dephasing (a) and photon time operation (b) protocols. (a) For the dressed dephasing sequence, a resonator probing tone of 400 ns is applied at different amplitudes with the qubit in superposition. The resonator probe frequency is swept for each amplitude. The 30 ns-long single qubit gates are $Y(+\pi/2)$ followed by $Y(-\pi/2)$ for measuring $<X>$ and  $Y(+\pi/2)$ followed by $X(+\pi/2)$ for measuring $<Y>$. The delay between the end of the resonator probe tone and the second qubit gate is 500 ns. (b) The photon time operation sequence starts with the resonator probe applying the measurement pulse used in our main experiments. A varying delay (150 ns in the Figure) is then applied, followed by a fixed-time echo experiment ($2\times 2.5 \mu$s).\label{fig:protocol}}
		
	\end{figure}
	
	\subsection*{Measurement characterization}
	
	In our cQED readout, each qubit is coupled to a co-planar waveguide (CPW) resonator, whose response to a microwave pulse depends on the qubit state. For the scope of this work we are only concerned with the measurement of the pointer qubit, whose details we discuss in this section.
	The CPW readout resonator is measured in reflection and the reflected signal is sent to a Traveling Wave Parametric Amplifier (TWPA) \cite{Yaakobi13}. Following the TWPA, the signal is further amplified at 3 K by a High Electron Mobility Transistor (HEMT) and then again at room temperature.
	The readout pulse is flat-top Gaussian shaped with ramp up and down $\sigma = 8$ ns and 300 ns total length. We calibrate the readout pulse frequency for maximum qubit state separation. The amplitude of the readout pulse is chosen so that the assignment error is as low as possible while staying well below the critical photon number \cite{Blais04}. For this particular readout, we find $n_{\mathrm{crit}}=\alpha \Delta/[4\chi (\Delta + \alpha)] \approx 23$. 
	\subsubsection*{Dressed dephasing experiments for measurement characterization}
	
	We present here a protocol for accurately measuring the readout cavity dispersive shift $\chi$ in the case where $\chi \lesssim \kappa$. This protocol presents advantages versus simpler two-tone spectroscopy approaches and time-domain ring-up and ring-down approaches in terms of accuracy and sensitivity. Some other advanced methods for this type of measurement use a more sensitive parameter estimation
	strategy, such as based on qubit dressed dephasing \citep{Schuster2005-ACStark,Gambetta2006-dephasing,Boissonneault2008,Gambetta2008-qm-traj,Boissonneault2009-Photon-induced-relax},
	to extract $\chi$. A use case is to sweep the cavity readout\textendash tone
	amplitude. From the sweeps, one extracts the qubit AC stark shift
	and induced dephasing. However, due to the highly non-linear response
	of the dephasing parameters as a function of $\chi$ and $\kappa$,
	this protocol is only most efficient in the case $\chi\ll\kappa$. 
	
	Our protocol schematic is visually
	depicted in Fig.~\ref{fig:protocol}(a). It consists of a Ramsey sequence modified with the additional cavity probe tone.
	The probe tone is swept in frequency over the range of the cavity
	peaks of interest. This process is valid down to single photon
	levels, and can be performed at $\bar{n}\ll1$. We repeat the measurements at several
	powers in order to increase our confidence in the extracted
	parameters. One further distinction versus previous protocols is that here both quadratures of the
	qubit, $\left\langle X\right\rangle $ and $\left\langle Y\right\rangle $, are measured, so that we can extract both the rotation and dephasing.
	
	\begin{figure}[H]
		\begin{centering}
			\includegraphics[width=\textwidth]{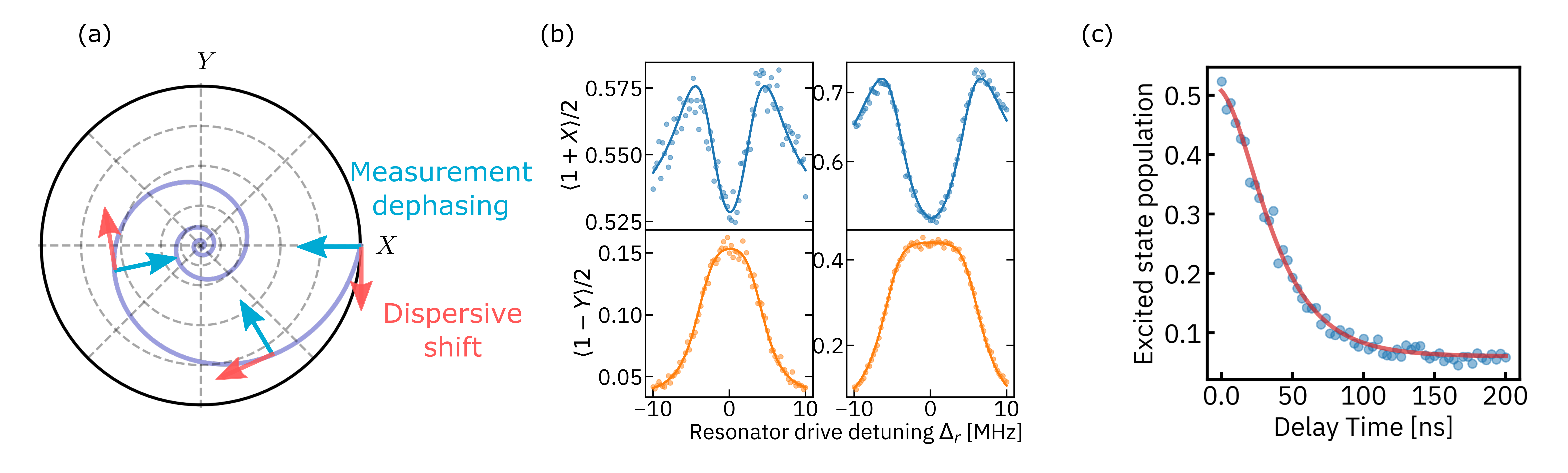}
			\par\end{centering}
		\caption{(a) Schematic depicting effect of
			measurement induced dephasing and dispersive stark shift on the qubit
			Bloch sphere equatorial plane. (b) Dephasing data with fits at 0.04 (left) and 0.28 (right) photons in the cavity. (c) Contrast decay, following Eq. \ref{eq:func4}, from the sequence depicted in \ref{fig:protocol}(b). The fit yields a number of photons for the measurement tone, after correcting for the finite qubit gate length, of $n_0 \sim 11$.  \label{fig:bloch}}
	\end{figure}

	We follow the discussion in Section III of Ref. \cite{Gambetta2008-qm-traj} to arrive at the equations governing the dynamics of the qubit during the measurement. This formalism eliminates the resonator degree of freedom from the full master equation by applying a polaron transformation. We can thus obtain the equations for the measurement-induced AC Stark shift and the measurement-induced dephasing
	
	\begin{equation}
	{\Omega_{ij}(t)=\left(\chi_{i}-\chi_{j}\right)\mathrm{Re}\left[\alpha_{i}^{*}(t)\alpha_{j}(t)\right]\,,}\label{eq:AC}
	\end{equation}
	
	\begin{equation}
	{\Gamma_{ij}(t)=\left(\chi_{i}-\chi_{j}\right)\mathrm{Im}\left[\alpha_{i}^{*}(t)\alpha_{j}(t)\right]\,,}\label{Dephase}
	\end{equation}
	where $\alpha_i$ is the readout resonator response to a finite duration square pulse drive of length $t_{\mathrm{P}}$ when the qubit is in state $|i\rangle$. 
	
	In order to integrate the above equations, we note that  $\alpha_i$ satisfies the first order differential equation:
	
	\[
	\dot{\alpha_i}(t)=a_{\mathrm{in}}\sqrt{\kappa}-\frac{1}{2}\kappa\alpha(t)-i(\Delta_r+\chi_i)\alpha(t)
	\]
	where $a_{\mathrm{in}}$ is the input operator for the resonator, $\kappa$ is the resonator decay rate, $\Delta_r$ is the detuning of the resonator drive and the resonator frequency, and $\chi_i$ is the dispersive coupling between the resonator and the qubit state $|i\rangle$. The above differential equation is subject to the boundary condition $\alpha_i (0) = 0$ and thus we can integrate the resonator response \textit{during} and \textit{after} the pulse, obtaining 
	\begin{align*}
	\alpha_{j}(t)= & -\frac{i \epsilon\left(-1+\exp\left(-t\left[\frac{\kappa}{2}+i(\Delta_r+\chi_{j})\right]\right)\right)}{\frac{\kappa}{2}+i(\Delta_r+\chi_{j})}\,,\\
	\alpha_{j}(t)= & -\alpha_i(t_{\mathrm{P}})\exp\left(-t\left[\frac{\kappa}{2}+i(\Delta_r+\chi_{j})\right]\right)\
	\end{align*}
	where we have used $a_{\mathrm{in}} = i\epsilon/\sqrt{\kappa}$ for a readout pulse amplitude $\epsilon$. Using Eqs.~\eqref{eq:AC} and \eqref{Dephase} and the Bloch equations
	\begin{align*}
	\dot{X} (t) = & -\omega_{\mathrm{ac}} (t) Y(t) - [\gamma_2 + \Gamma_{ge} (t)] X(t) \\
	\dot{Y} (t) = & \, \omega_{\mathrm{ac}} (t) X(t) - [\gamma_2 + \Gamma_{ge} (t)] Y(t) 
	\end{align*}
	where $\omega_{\mathrm{ac}} (t) = \omega_{\mathrm{q}} + \Omega_{ge} (t)$, $\omega_{\mathrm{q}}$ is the Lamb-shifted qubit frequency, and $\gamma_2$ is the qubit dephasing rate, we can numerically fit experimental data as a function of the readout drive detuning. The canonical action of the measurement-induced dephasing and dispersive shifts are shown in Fig. \ref{fig:bloch}(a).
	
	In Fig. \ref{fig:bloch} (b) we show the expectation values of $1+X$ and $1-Y$ as a function of $\Delta_r$ for DAC amplitudes of 0.01 V (left) and 0.025 V (right). From these fits we obtain a dispersive shift value of $\chi/\pi = 5.8$ MHz and a cavity decay time of $\kappa/2\pi = 5.7$ MHz. Fitting the average photon number as a function of the drive power to a linear function [Fig. \ref{fig:bloch}(c)], we can extrapolate to the measurement drive power used in our algorithm experiments to obtain an average photon number of $\bar{n} \sim 11$.
	
	These are highly non-linear curves which are superbly sensitive to $\chi$, $\kappa$, and $\bar{n}$. All
	three parameters can be calibrated simultaneously, yielding more information than most other protocols. 
	Two aspects of this method contribute greatly to its sensitivenes: first, the distinguishability in the cavity conditional response can be understood as vector operations, such as the cross-product, which contain the parameters in both a numerator and a denominator; and second, using the qubit to perform a Ramsey measurement provides the standard interferometer advantage and insensitivity to offsets and scaling biases in the data. 
	
	\subsubsection*{Measuring readout photon number via photon time operation}
	The sequence for measuring the readout resonator photon population at the end of the readout pulse is inspired in Ref. \cite{McClure16}. The functional form of the echo experiment depicted in Fig. \ref{fig:protocol}(b) is
	\begin{equation}\label{eq:func}
	S = \frac{1}{2}[1-e^{-\Gamma_2 t_R} \mathrm{Im} (\exp(-(\phi_0 + 2n \chi \tau)i))]
	\end{equation}
	which follows from Eq. (1) in Ref. \cite{McClure16} particularized for zero Ramsey detuning $\Delta_R = 0$. Here, $\Gamma_2$ is the dephasing rate of the qubit, $\phi_0$ is the initial phase, $n$ is the number of resonator photons, $\chi$ is the qubit-resonator energy coupling, and $\tau = (1 - e^{-(\kappa + 2 \chi i)t_{\mathrm{R}}})/(\kappa + 2 \chi i)$, with $\kappa$ the resonator decay rate and $t_R$ the Ramsey delay [5 $\mu$s in our case, see Fig. \ref{fig:protocol}(b)]
	
	We choose to apply an echo to the Ramsey experiment to remove the contribution from external low-frequency noise. This readout photon dynamics experiment carries some model assumptions that have to be met for a trusted fit:
	\begin{itemize}
		\item $\kappa t_{\mathrm{R}} \gg 1$ This condition removes any significant time dependence in $\tau$ which would result in a reduced measurement induced dephasing
		\item $1/t_g \gg n \kappa$ We assume that the length of the qubit gates used for the echo pulses is much shorter than all other dynamics involved in the system. Longer gates will result in systematic offsets at zero delay, which may translate in an underestimate of the photon number $n_0$. In our case, for $\kappa/2\pi = 5.7$ MHz and $t_g = 30$ ns, we are not quite in that regime. Therefore we can assume that, at zeroth order, we can rescale the obtained $n_0$ by a factor $e^{\kappa t_g} \sim 2.928$.
		\item $1/\Delta t_{\mathrm{timing}} \gg n_0 \kappa$ where $\Delta t_{\mathrm{timing}}$ takes into account relative delay in the signal path of the qubit and resonator probe tones. In our case, we calibrate that timing down to the nanosecond level.
	\end{itemize}
	
	Applying the first condition, $\kappa t_{\mathrm{R}} \gg 1$, we can write $\tau = 1 / (\kappa + 2 \chi i)$. Equation \ref{eq:func} then becomes
	\begin{equation}\label{eq:func2}
	\begin{split}
	S = \frac{1}{2}\Big[1-e^{-\Gamma_2 t_R} \exp(-\frac{4n \chi^2}{\kappa^2 + 4\chi^2})\mathrm{Im} \Big(\exp\Big(-(\phi_0 + \frac{2n\chi\kappa}{\kappa^2 + 4\chi^2})i\Big)\Big)\Big]\\
	= \frac{1}{2}\Big[1-e^{-\Gamma_2 t_R} \exp(-\frac{4n \chi^2}{\kappa^2 + 4\chi^2})\sin \Big(-(\phi_0 + \frac{2n\chi\kappa}{\kappa^2 + 4\chi^2})\Big)\Big]
	\end{split}
	\end{equation}
	An echo sequence with no dephasing should yield $S=0$, so from Eq. \ref{eq:func}, and setting $n = \Delta_R = \Gamma_2 = 0$, we obtain $\phi_0 = -\pi/2$. Substituting in Eq. \ref{eq:func2}:
	\begin{equation}\label{eq:func3}
	S = \frac{1}{2}\Big[1-e^{-\Gamma_2 t_R} \exp(-\frac{4n \chi^2}{\kappa^2 + 4\chi^2})\cos \Big(\frac{2n \chi \kappa}{\kappa^2 + 4 \chi^2}\Big)\Big]
	\end{equation}
	
	We can now define the quantities $\alpha = e^{-\Gamma_2 t_{\mathrm{R}}}$, which gives us a measure of the dephasing, and $\beta = 4\chi^2 / (\kappa^2 + 4 \chi^2)$, which gives us a measure of the distinguishability. In particular, when driving the cavity halfway between the resonances for the qubit in the ground and excited states, one gets
	
	\begin{equation*}
	D = |\alpha - (-\alpha)|^2 = 4|\alpha|^2 = \frac{4n_0 \chi^2}{\kappa^2/4 + \chi^2} = 4 \beta n_0
	\end{equation*}
	
	Now, as a function of the delay time in Fig. \ref{fig:protocol}(b) $t$ we can express the functional form in Eq. \ref{eq:func3} as follows
	\begin{equation}\label{eq:func4}
	S = \frac{1}{2}\Big[1-\alpha \exp(-\beta n_0 e^{-\kappa t})\cos \Big(\beta n_0 e^{-\kappa t} \frac{\kappa}{2\chi}\Big)\Big]
	\end{equation}
	
	where the number of photons in the resonator as a function of time has been written as $n_0 e^{-\kappa t}$. The cosine term of the equation shifts the qubit frequency depending on the photon population of the resonator. Simultaneously, there is an exponential loss of contrast as the resonator depopulates. The experimental results of this experiment and the fit to Eq. \ref{eq:func4} are shown in Fig. \ref{fig:bloch}(c). We enforce the $\chi$ and $\kappa$ values obtained from the dressed dephasing protocol and leave only $\alpha$ and $n_0$ as free parameters in Eq. \ref{eq:func4}. From the fit, we obtain $n_0 \sim 3.8$. Applying the $e^{\kappa t_g} \sim 2.93$ correction discussed above, we find $n_0 \sim 11$, which is about half of $n_{\mathrm{crit}}$

	\subsection*{Overview of the control electronics}
	
\begin{figure}[H]
	\centering
	\includegraphics[width=\textwidth, keepaspectratio]{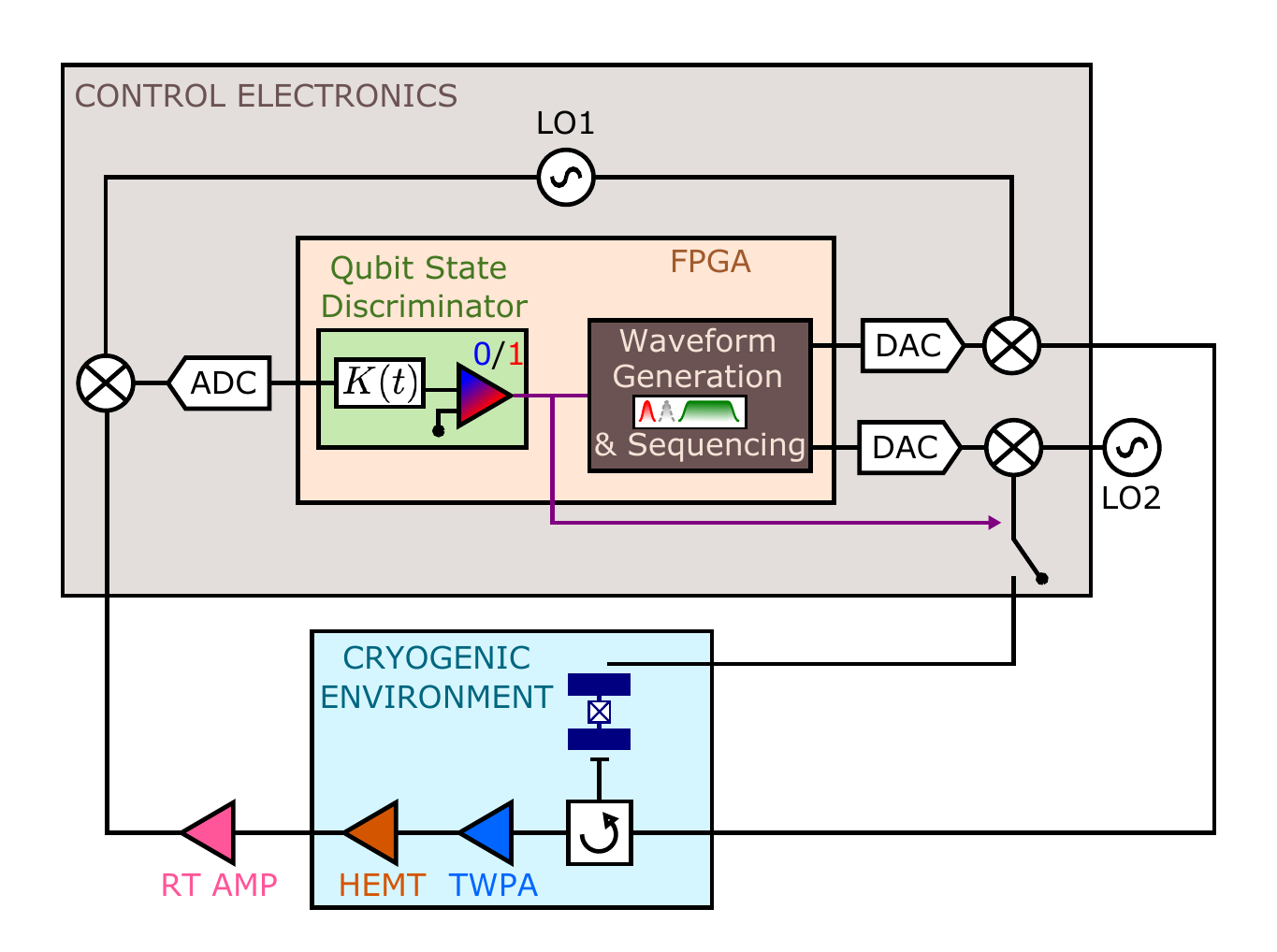}
	\caption{\label{SLICE} Schematic of the quantum computing system including quantum and classical hardware.}
\end{figure}

	The control electronics provide both Digital-to-Analog (DAC) conversion for qubit and measurement tone production and Analog-to-Digital (ADC) conversion for measurement signal acquisition and analysis. Fig. \ref{SLICE} shows a schematic of the control electronics and the cryogenic system. Prior to the DAC, a waveform generation and sequencing module within the FPGA is used to define the pulses amplitudes, intermediate frequency (IF) , shapes and durations and to orchestrate their playing sequence. Both the qubit control and readout pulses are upconverted by mixing with a Local Oscillator (LO). The control electronics include an RF synthesizer that provides the readout LO (LO1) whereas an external LO (LO2) is used for qubit control pulses. The readout resonator is probed in reflection and the outcoming reflected signal is amplified by a Traveling Wave Parametric Amplifier (TWPA) \cite{Macklin15} at the coldest refrigerator stage. Further amplification is applied at 3 K (HEMT) and at room temperature.
	
	The high speed low latency classical control electronics for QPE experiments evolved from a prototype \cite{Bronn17} 
	that comprised an FPGA card, ADC and DAC cards, 
	along with commercial off the shelf radio frequency instruments and discrete components 
	such as RF signal generators, switches and amplifiers.  
	The prototype has been extensively re-engineered as a set of integrated custom cards 
	to support RF waveform generation and readout capture 
	under host PC control via a PCI Express (PCIe) Gen 2 / Gen 3 x4 link. 
	
	The current generation of control electronics family supports two DAC channels 
	that run up to 2.5 GS/s (2.5e9 samples per second) at up to 14 bit resolution for arbitrary waveform generation (AWG) functionality 
	to support qubit control and readout, as well as two ADC channels of up to 500 MS/s (5.0e8 Samples per second) at up to 14 bit resolution 
	for qubit state determination, with an optional provision to configure the ADC for 1 GS/s sampling rate single channel. EV12AS2000A and EV12DS4000A chips were used in the experiments described in this work.
	
	A Xilinx MPSoC UltraScale+ FPGA XCZU19EG drives the ADC and DAC channels 
	with a custom programmable logic, designed specifically for fast qubit state determination and 
	conditional playout of control and readout waveform sequences. 
	
	A clock distribution network comprising of several phase locked loop (PLL) clock generator chips provides 
	flexible ADC, DAC and readout local oscillator (LO) clock generation with very low jitters. 
	A 4GB DDR4 memory module, designed to accommodate up to 32GB, and an Ethernet port 
	are provided for the embedded ARM microcontroller in the Xilinx FPGA (currently unused) for potential future enhancements. 
	
	We describe the ADC controller FPGA logic and the DAC controller FPGA logic in the following sections. 
	
	\subsubsection{ADC Controller Logic}
	To support qubit state readout, the ADC chip captures and digitizes down-converted intermediate frequency signals at 1G samples per second, 
	which is then transmitted to the ADC controller logic in the FPGA via a set of parallel low voltage differential signaling (LVDS) pairs 
	with a 1:2 multiplexing ratio, interleaved into two physical channels. 
	Each of the two channels supports 500MS/s sampling rate with a phase offset of 1 ns, accommodating up to 128K samples. 
	Only one channel is enabled for the QPE experiments because the intermediate frequency was kept fairly low, 
	in the 25-50MHz range, thanks to use of individually configurable readout LO generator. 

	A serializer / de-serializer (SERDES) I/O block in the Xilinx FPGA was then used in 4-to-1 mode 
	to receive digitized data from the ADC chip. 
	
	The ADC controller logic implemented in the FPGA supports high bandwidth DMA data transfer over the PCIe link to the host system 
	with an effective sustained transfer rate of up to 1.6GBytes/sec depending on transfer block size, 
	which is more than sufficient for typical experiments 
	that measures qubits at 1 ms trigger interval to transfer the whole ADC capture sample memory to the host PC without losing data. 
	
	To complement the high bandwidth, the ADC controller also implements an optional ``trigger aggregation'' feature, 
	in which the FPGA logic accumulates multiple triggers' worth of data (up to 256 triggers) on FPGA memory in a ring buffer format. 
	This feature was developed for experiments that require bursts of fast-repeating, relatively short readouts, 
	allowing the captured data blocks to be transferred in batches rather than one block at a time on each readout trigger. 
	This minimizes the impact of PCIe transaction overhead, which can be on the order of microseconds per DMA transfer.
	
	The ADC sample memory is implemented by using `dual port' type of FPGA memory, 
	which allows the DMA data transfer to host to occur asynchronously and in parallel with the qubit readout operation, 
	so that there will be no data loss 
	as long as the aggregate readout data rate (readouts per second multiplied by readout sample length) 
	does not exceed the effective transfer rate. 
	
	The ADC controller implements a matched filter \cite{Ryan:2015} in FPGA logic 
	in order to determine qubit measurement results with a very low latency and high fidelity. 
	We first calculate the appropriate kernel weight function and the associated threshold value 
	based on known $|0 \rangle$ and $|1 \rangle$ calibration runs, load the results in FPGA memory, 
	then integrate the incoming readout data stream with the kernel at runtime. 
	The integration result is compared against the threshold value derived from the calibration runs 
	to generate a binary readout result of 0 or 1 representing the measured qubit state, as illustrated in Fig. \ref{SLICE}.
	
	The run-time processing on FPGA is done by using Xilinx digital signal processing (DSP) blocks built into the FPGA, 
	which yields a combined overhead of up to four logic clock cycles for integration and thresholding. 
	Experimentally we have found this design to be easier to optimize to the same or higher fidelity 
	compared to conventional intermediate frequency (IF) demodulation methods that use digital down-conversion 
	via mixing with an IF carrier and low pass filtering in software or hardware. 
	The design also generated the readout results with less latency for the maximum fidelity 
	on the qubit devices used in the QPE experiments. 

	The readout determination results are brought out to logic output ports to control an analog RF switch 
	that gates qubit reset pulses for active reset, and/or 
	to drive the conditional branch logic in the FPGA DAC controller (described in the next section). 
	The nominal overall latency from ADC trigger to readout result, excluding the readout data stream sample size 
	(which depends on the qubit), is 40 ns in the QPE experiment configuration. The main body of this work describes the systemic latencies in more detail.
	
	The determination results are also made available to the host software along with the raw data, 
	for cross validation as well as to provide for extreme scaling in the near future, 
	when aggressive offloading of data processing from host software to FPGA hardware would be desired 
	in order to support hundreds of qubits in a system.
	
	\subsubsection{DAC Controller Logic}
	
	For waveform generation, the FPGA DAC controller logic supports two operating modes: 
	
	1) a Sequence Processor (SP) mode that implements a 128-bit Very Long Instruction Word (VLIW) architecture 
	designed to support waveform re-use and loops (for reduced memory footprint), 
	absolute branches and conditional branches on external control signal 
	(typically the readout result signals generated by the FPGA ADC controller), 
	and sideband signals (markers) to control external analog gates and secondary triggers including ADC trigger; and 
	
	2) a simpler operating mode supporting single waveform and marker data block per trigger, controlled by a number of software managed registers.
	
	The SP mode was used in the QPE experiments described in this work. The 10 bit IPE experiments used up approximately 30 percent of the available 64K instructions.
	
	The SP VLIW control logic in the DAC data flow path 
	operates synchronous to the data transfer clock from the DAC chip, whose frequency is configured to 1/8th of the DAC sampling clock.   
	Clocking at this speed has allowed the logic design to implement single clock cycle conditional branching 
	based on results of qubit state readout. 

	A 128bit wide FPGA block RAM (BRAM) is used to store the waveform memory, accommodating up to 512K samples, so that 8 samples are fetched in parallel. 
	A Xilinx FPGA SERDES runing in 8-to-1 mode is then used to send the sample values 
	over the LVDS parallel link to the two DAC chips.  
	
	The SP instruction format is illustrated in Table \ref{table:fpga1}: 
	
	\begin{table}[h!]
		\begin{center}
			\caption{Sequence Processor Instruction Format}
			\label{table:fpga1}
			\begin{tabular}{|l|l|l|}
				\hline
				\textbf{Field} & \textbf{Bits} & \textbf{Description}\\
				\hline
				Sample Offset & 24 & Starting offset of output samples\\
				\hline
				Sample Count & 24 & Number of output samples\\
				\hline
				Loop Count & 10 & Repeat count \\ 
				\hline
				Branch Index 1 & 16 & Branch target 1 \\
				\hline
				Branch Index 2 & 16 & Branch target 2 \\
				\hline
				Control & 10 & Execution control including: \\
				&   & - Trigger wait \\
				&   & - Next sequential instruction / \\
				&   & \ \ Unconditional branch / \\
				&   & \ \ Conditional branch \\
				\hline                  
			\end{tabular}
		\end{center}
\end{table}
	
    To describe quantum programs that take advantage of the sequence processor, a prototype extension
    to existing quantum programming language has been developed to include labels, branches, and gotos.
    The branch primitive causes a quantum gate sequence to branch conditionally to a target location
    identified by a label, while the goto causes an unconditional branch. 
    
    The follow pseudo-code illustrates how these primitives are used in a conditional reset example.
    \newpage
    \begin{verbatim}
    start:
      rx(pi / 2) q1;
      measure q1 -> c;
      bnz c, A;
      id q1;
      goto B;
    A:
      x q1;
    B:
      halt;
    \end{verbatim}
    
    At the start of the sequence, the qubit \texttt{q1} receives a $\pi/2$-pulse and then the result of
    a measurement is stored in the register \texttt{c}. When the register is non-zero (the qubit is in
    the $|1\rangle$ state), we jump to label \texttt{A} to flip the qubit back to $|0\rangle$ with an X
    gate. When the register is zero, we insert an identity gate to ensure that the pulse sequences
    maintain the same phase regardless of which path is taken. (This part is not executed in the actual
    experiment, and is optimized out during pulse generation.) In this example, the two conditional
    paths converge at label \texttt{B}.

	The main body of the QPE algorithm, which shifts the phase of the control signal based on previous measurement results, 
	has been implemented by using this extension. 
	
	This extension could also be used to implement qubit reset that is needed after each measurement between blocks of controlled-U operations in the QPE algorithm. 
	However, in the QPE experiment we chose to implement the qubit reset operations 
	by using an integrated RF switch that is driven by the readout result signal to conditionally gate the qubit reset pulse. 
	This was done to reduce the required amount of instruction memory and the complexity of quantum software. 
	Likewise, although the FPGA DAC controller logic and the DAC chips support sampling rates of up to 2.5 GS/s, 
	the sampling rate was kept down to 2 GS/s using the programmable clock generation network 
	for ease of quantum program development. 
	
	The software framework for the FPGA DAC controller (waveform data generator and player) is 
	an extension of the existing IBM Quantum software backend, 
	with hardware specific accommodations and enhancements to support conditional branches. 
	
	\subsection*{Reset fidelity and readout QND-ness}
	We showed in the main text a brief demonstration of qubit measurement and reset where we prepared alternate ground and excited qubit states, read out the qubit, and conditionally reset it at repetition rates much faster than the qubit lifetime. This gave a sense for the amount of state preparation error present in our measurement and reset apparatus as compared to qubit environmental thermalization. In this section we want to explore this tool a bit deeper and try to gauge the amount of qubit measurement-induced back-action present in our system by quantifying the  reset fidelity and its dependence on the length of a sequence of measurement and reset steps. We show the circuit used in the bottom panel of Fig. \ref{fig:ipe_rep_meas}(b). It consists of a Hadamard gate followed by four cycles of measurement and conditional reset. This sequence is repeated at a low rate of 1 kHz, allowing the qubit to thermalize before the next Hadamard.
	
	For the definition of a reset error metric we use the Hellinger distance, which gives the distance between two probability distributions. In the case of two discrete distributions $P = (p_1, p_2, \dots , p_k)$ and $Q = (q_1, q_2, \dots , q_k)$ the Hellinger distance $H(P,Q)$ is defined as 
	\begin{equation*}
	H^2(P,Q) = \frac{1}{2} \sum_{i=1}^k (\sqrt{p_i} - \sqrt{q_i})^2
	\end{equation*}
	If $Q$ is the ideal distribution, one can define the fidelity associated to $P$ as the square of the Bhattacharya coefficient, $BC^2 = (1-H^2(P,Q))^2$. In the case of a reset protocol, we have $Q = (1,0)$ and $P=(P(0), P(1))$, which results in $H^2(P,Q) = (1 - \sqrt{1-P(1)})$. The reset error can thus be defined as the probability of obtaining the exicted state $P(1)$.
	
	We bin (and later histogram, Fig. \ref{fig:ipe_rep_meas}(b) top panel) the measurement result and apply an $X$ gate conditional to the qubit read out in the excited state (green dashed boxes). For each measurement we show the probability of measuring 0, $P(0)$, as well as the reset error, $P(1)$. We find a reset error of $1.65\%$ after one reset and of $1\%$ after two resets. A subsequent iteration of measurement and reset does not significantly increase the reset error, so we can trust our measurement is reasonably Quantum Non-Demolition (QND). For the main experiments in this work we used two cycles of measurement and reset for the pointer qubit.
	
	\subsection*{IPE Algorithm output from circuit statistics}
	The standard textbook version of quantum phase estimation \cite{nielsen00,Cleve98} has a probability of finding the answer within an error of $1/2^{m+1}$ for an $m-$bit implementation of the algorithm which is higher than the constant $8/\pi^2$ \cite{Kaye07}  with $m$ measurements \textit{independent of} $m$. When many measurements per bit are allowed, however, the question of what is the best approximation to the answer from the sampled output distribution of the algorithm circuit or circuits naturally arises. Given that the answer will always lie between two of the possible output bitstrings, which, if the noise is low enough, should be sampled with higher frequency than the rest of the outcomes, it seems natural to define the algorithm output as the weighted average of these two bitstrings. Due to sampling noise, however, simply taking the average of the two most sampled bitstrings may not give the lowest error. One can however estimate the phase as the weighted average of the two most likely \textit{consecutive} bitstrings. 
	
\begin{figure}
		\begin{centering}
			\includegraphics[width=\textwidth]{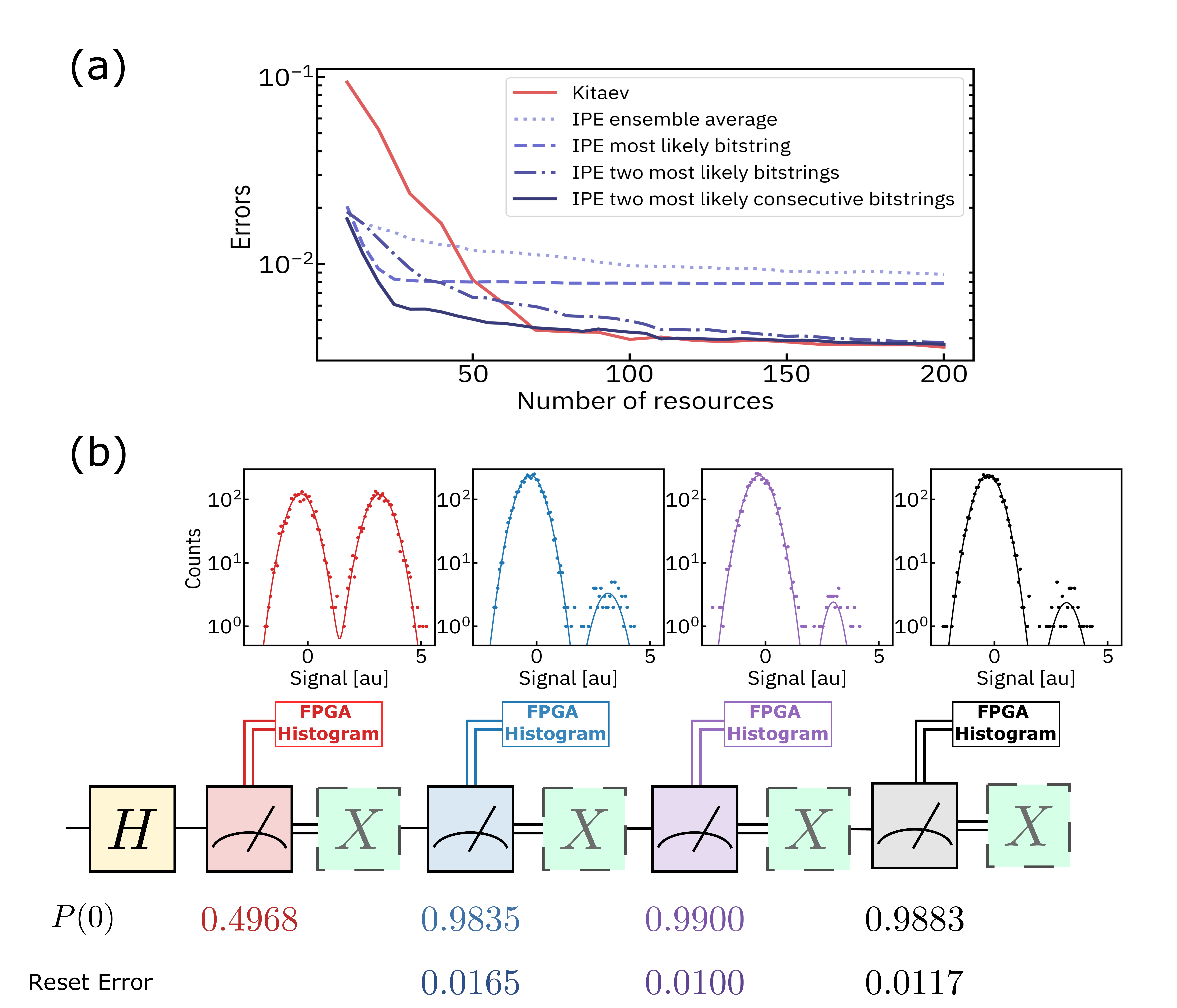}
			\par\end{centering}
		\caption{(a) Noiseless simulation of both quantum phase estimation methods presented in this work for 5 bits accuracy. The IPE answer is computed from the output distribution of the circuits in four different ways: by computing the total ensemble average, by selecting always the most likely bitstring, and by computing the weighted average of the two most likely bitstrings with and without imposing the constrain that they be consecutive. We find that the most accurate result is the weighted average of the two most likely consecutive bitstrings. We also show the output of the Kitaev protocol in red for comparison. (b) Demonstration of repeated measurement and reset on the pointer qubit. The qubit is initialized in the $|+\rangle$ state and then measured and reset four consecutive times. Two consecutive cycles of measurement and reset perform better than one, with subsequent repetitions not adding benefit.   \label{fig:ipe_rep_meas}}
	\end{figure}

	In Fig. \ref{fig:ipe_rep_meas}(a) we show a noiseless simulation for 5 bits with different solutions calculated from the output circuits distributions. We also show the Kitaev solution for comparison. We clearly see that simply taking the most likely bitstring soon flattens out as we sample the circuits outputs. The ensemble average includes the entirety of the sampling error and thus is the highest error of the four computations. At low number of resources, the weighted average of the two most likely bitstrings includes a significant amount of sampling error, which we can avoid by forcing the additional condition that the two most likely bitstring are consecutive. This latter algorithmic output consistently yields the lowest error at for any number of resources allowed and is the one we have used for the main body of this work. For comparison, we see that the Kitaev version of the algorithm has remarkably poor performance for very low number of resources whereas it tends to do equally well as the iterative version when the measurements are significantly increased.
	
	\subsection*{Kitaev Estimator}
	As noted in the main text, Kitaev's original approach to finding the eigenvalues of a unitary matrix offers an exponential advantage versus classical methods. However, an estimator needs to be built to process the outputs of the Kitaev quantum circuits and to find the approximation to the phase iteratively bit by bit. We show here how to construct such estimator.
	
	The circuits in the top panel of Fig. 1 in the main text provide estimations of the \textit{shifted} bits $\alpha_k = 2^{k-1}\tilde{\varphi}$, where $\tilde{\varphi}$ is the $m$-bit approximation to $\varphi$, $\tilde{\varphi}=\sum_{k=1}^m \varphi_k / 2^k = 0.\varphi_1 \varphi_2 \dots \varphi_m$ with $\varphi_k \in \{0,1\}$. Once we obtain all the $\alpha_k$ for $k$ from 1 to $m$, we can retrieve $\tilde{\varphi}$ using the following algorithm.
	
	\begin{algorithm}[H]
		\caption{Kitaev Estimator}
		\begin{algorithmic}[1]
			\State Estimate all the $\alpha_k$ using the circuits in Fig. 1b in the main text using the maximum \textit{total} allowed of measurement resources among the two circuits
			\State Set $0.\varphi_m \varphi_{m+1} \varphi_{m+2} = \beta_m$ where $\beta_m$ is the closest octant $\{\frac{0}{8}, \frac{1}{8}, \dots, \frac{7}{8} \}$ to $\alpha_m$
			\For{\texttt{$j=m-1$ to $1$ }}
			\State $\varphi_j = \begin{cases}
			0 \, \, \text{if } |0.0\varphi_{j+1}\varphi_{j+2} - \alpha_j|_{\mod 1} < 1/4 \\
			1  \, \, \text{if } |0.1\varphi_{j+1}\varphi_{j+2} - \alpha_j|_{\mod 1} < 1/4
			\end{cases}$
			\EndFor
			\State{\Return $\overset{\approx}{\varphi} = 0.\varphi_1 \varphi_2 \dots \varphi_{m+2}$}, the $(m+2)$-bit approximation to the phase $\varphi$.
		\end{algorithmic}
	\end{algorithm}
	
	As an example, consider the phase problem $\varphi=0.4840845$. We want to find the best 5-bit approximation. After 100 shots we find that $\alpha_5 = 2^4 \tilde{\varphi} = 0.7373$, which sets $\varphi_5 \varphi_6 \varphi_7 = 110$, since $|\alpha_5 - 0.110| < |\alpha_5 - 0.101|$ (where we are using the binary expansion notation introduced above). Note that even at the lowest possible number of resources, one shot per physical measurement, the angle obtained from both Kitaev circuits can only be, in units of $2\pi$, 0.125, 0.375, 0.625, or 0.875, which has a precision of three bits, thus expanding the least significant bit into three. We next measure $\alpha_4 = 0.8530$, which sets $\varphi_4 = 1$, since $|\alpha_4 - 0.111| < 1/4$ and $|\alpha_4- 0.011| > 1/4$. Continuing with this procedure, we get $\varphi_3=1$ from $\alpha_3 = 0.9208$, $\varphi_2 = 1$ from $\alpha_2 = 0.9649$, and $\varphi_1 = 0$ from $\alpha_1=0.4748$. We thus reach a 7-bit approximation to $\varphi$, $\overset{\approx}{\varphi} = 0.0111110 = 0.484375$, which gives an error to the phase problem $\varphi=0.4840845$ of $2.905 \times 10^{-4} < 1/2^8$. 
	
	\bibliographystyle{unsrt}
	\bibliography{Suppbib}